\documentclass[english,secnumarabic,amssymb, nobibnotes, aps, pre, preprint]{revtex4-1}
\usepackage[T1]{fontenc}
\usepackage{amsmath}
\usepackage{amssymb}
\usepackage{graphicx}
\usepackage{esint}
\usepackage{bibunits}
\usepackage{tikz}
\usetikzlibrary{chains}
\usetikzlibrary{arrows}

\makeatletter

\providecommand{\tabularnewline}{\\}
\newcommand{\lyxdot}{.}

\@ifundefined{textcolor}{}
{%
 \definecolor{BLACK}{gray}{0}
 \definecolor{WHITE}{gray}{1}
 \definecolor{RED}{rgb}{1,0,0}
 \definecolor{GREEN}{rgb}{0,1,0}
 \definecolor{BLUE}{rgb}{0,0,1}
 \definecolor{CYAN}{cmyk}{1,0,0,0}
 \definecolor{MAGENTA}{cmyk}{0,1,0,0}
 \definecolor{YELLOW}{cmyk}{0,0,1,0}
}

\newcommand*{\rom}[1]{\expandafter\@slowromancap\romannumeral #1@}

\@ifundefined{showcaptionsetup}{}{%
 \PassOptionsToPackage{caption=false}{subfig}}
\usepackage{subfig}
\makeatother

\usepackage{letltxmacro}

\LetLtxMacro{\ORIGselectlanguage}{\selectlanguage}
\makeatletter
\DeclareRobustCommand{\selectlanguage}[1]{%
  \@ifundefined{alias@\string#1}
    {\ORIGselectlanguage{#1}}
    {\begingroup\edef\x{\endgroup
       \noexpand\ORIGselectlanguage{\@nameuse{alias@#1}}}\x}%
}
\newcommand{\definelanguagealias}[2]{%
  \@namedef{alias@#1}{#2}%
}
\makeatother

\usepackage[english]{babel}
\definelanguagealias{en}{english}
\begin{document}
\begin{bibunit}

\title{Does Having More Options Mean Harder to Reach Consensus? (Accelerating
Consensus By Having More Options)}

\author{Degang Wu, Kwok Yip Szeto}
\begin{abstract}
We generalize a binary majority-vote model on adaptive networks to
a plurality-vote counterpart. When opinions are uniformly distributed
in the population of voters in the initial state, it is found that
having more available opinions in the initial state actually accelerate
the time to consensus. In particular, we investigate the three-state
plurality-vote model. While time to consensus in two state model scales
exponentially with population size $N$, for finite-size system, there
is a non-zero probability that either the population reaches the consensus
state in a time that is very short and independent of $N$ (in the
heterophily regime), or in a time that scales exponentially with $N$
but is still much faster than two-state model. 
\end{abstract}
\maketitle

\section{Introduction}

Interest in problems of voting dynamics and opinion formation are not limited to social-political studies, as many models constructed by physicists and mathematicians have been designed to estimate the time needed to reach consensus. Examples are the voter model~\cite{clifford_model_1973, holley_ergodic_1975}, majority-rule model~\cite{galam_minority_2002}, Sznaj model~\cite{katarzyna_opinion_2000,sabatelli_non-monotonic_2004}, Axelrod's model~\cite{axelrod_dissemination_1997,castellano_statistical_2009} etc. For a review of major models refer to \cite{sen_sociophysics:_2013}.
One of the key question concerns the evolution of opinion in a multi-agent system, where the agents can be modelled by ``particles'' with special attribute and interactions that can also be changing with time. The agents, voters, or particles are modelled as nodes in a social network, with links between nodes specifying their interaction. Since a changed opinion (or attribute) of the agent can induce change to the connection with the neighboring nodes, while a changed connection can also induce a change to the opinion of the agent, the entire system of interacting agents is therefore a co-evolving networks  with both nodes and links changing. The goal of opinion formation is to count the number of agents holding a particular opinion as a function of time, but the fact that the links connecting nodes are also changing with time implies that we are addressing a problem of great complexity defined on a ``social network'' of evolving topology. 
The complexity of this problem is further accentuated by the deadline imposed on the specific election.  
Consequently, the usual studies of time scale to reach consensus in voting model must be rephrased in terms of the speed to consensus. 
For example, a party in an election may win in the long run, but in the short run, such as at the deadline for counting the vote, another party may have more votes and end up winning. Therefore a comparison of time scales for the opinion formation process is very important in application. 
In this paper, we address this question of time scales from the perspective of three-state plurality-vote model. 

While agent-based simulations were frequently employed to study co-evolving opinion dynamics, the extension to large scale usually encounter problems due to the complexity of the model with updating rules that are complex if the model is realistic~\cite{demirel_moment-closure_2014}. Therefore a complementary approach is to build simpler, but mathematically amenable models such as the one proposed by Benczik et al.~\cite{benczik_opinion_2009}, so as to extract valuable insights to understand the qualitative behaviors observed in simulation. The opinion in these mathematical models can either be a discrete~\cite{galam_minority_2002,katarzyna_opinion_2000,de_oliveira_isotropic_1992}, or continuous variable \cite{hegselmann_opinion_2002}, while the exact interpretation of the opinion is very flexible, depending on the context in application. For example, 
opinions could be political views to adhere to, sports teams to support, musical styles people enjoy, and so on. For the discrete models, most research focus on the simplest two-state model, i.e., the opinion is ``yes'' or ``no'' response to an issue. 
Recent work suggests that the time to consensus increases with the number of available opinions~\cite{wang_freezing_2014}, while other numerical work~\cite{wu_three-state_2014} suggests otherwise. 
Since the nature of the increase or decrease in time to consensus is still unclear, we like to clarify this issue for the case of three-opinion model. The conclusion of our investigation must also be tested for large population, so that scaling behavior of the time to consensus must be addressed, Our starting point is to generalize the binary majority-vote model on adaptive networks~\cite{benczik_opinion_2009} to plurality-vote model with more than two states. 
Our approach is mainly numerical, but we also use analytical results to verify our numerical results to achieve a better understanding of  the mechanism behind the various time scales to consensus derive their scaling relation with the population size $N$.  
In different context, Refs.~\cite{gekle_opinion_2005, galam_drastic_2013} investigated three opinion system with discussion-group-dynamics.
The focus was on the dominance of minority opinion due to hidden preferences in case of a tie in voting and the size of discussion group is fixed so as to allow full analytical treatment.
Our approach is mainly numerical, but we also use analytical results to verify our numerical results to achieve a better understanding of  the mechanism behind the various time scales to consensus derive their scaling relation with the population size $N$.  
In
Sec.~\ref{sec:Model}, we introduce the plurality-vote model on adaptive
networks. In Sec.~\ref{sec:Simulation}, we present the acceleration
of consensus induced by having more than two states by simulation
results. In Sec.~\ref{sec:M-Equation}, the M-equation, or master
equation, for the plurality-vote model is derived and analyzed. The
mechanism behind the acceleration of consensus will be examined in
Sec.~\ref{sec:Mechanism}. Finally, concluding remarks will be presented
in Sec.~\ref{sec:Conclusions}.

\section{Model\label{sec:Model}}

Our model consists of $N$ agents (nodes), each carries an opinion
$\sigma_{j}=1,2$ or $3$, with $j=1,2,\dots,N$. Agents and links
coevolve according to the following dynamics. 
In each time step, we randomly choose an agent $i$ to be updated. 
Temporary links will be formed between $i$ and other agents in the population, according to a probability $p$ and $q$, which are constants among the whole population.
We go through all possible edges between $i$ and $j$, where $j=1,2,\dots N$
and $j\neq i$. If $\sigma_{i}=\sigma_{j}$, then a link will be formed
between the two nodes with a probability $p$. If $\sigma_{i}\neq\sigma_{j}$,
a link will be formed with a probability $q$. Here we assume $q\equiv1-p$. 
Once we have decided all the temporary links between agent $i$ and
all other agents, we update $i$ using the following rule: we count
the number of the three opinions in $i$'s temporary neighborhood.
If there is a plurality opinion in the temporary neighbors $(v)$, then we update the agent $i$'s opinion by $\sigma_{i}=v$; otherwise $\sigma_{i}$ remains unchanged.
Here, by \emph{plurality}, we mean the situation when the number of one opinion is larger than the number of any of the other opinions.
Therefore, in this work, majority is a special case of plurality.
This update rule is very similar to the majority rule model~\cite{galam_minority_2002}.
After the update, all temporary links are eliminated. The temporary
nature of the link formation process renders our model amenable to
mean-field like mathematical treatment. 
The structure of our model is similar to the two-opinion model of Benczik et al.~\cite{benczik_opinion_2009}, so that the temporary nature of the link formation renders our model amenable to mean field analysis. 

In our model, large $p$ or small $q$ could indicate that individuals
are more likely to hear from people holding the same opinion (homophily)
or supporting the same political party. Small $p$ or large $q$ may
represent the situation where individuals are more likely to interact
with people with different and diverse background (heterophily) or
not satisfied with the original opinion or party, and are seeking
for a different opinion. However, our model does not assume that the
voters are homophily or heterophily in their nature. In fact, we can
have other interpretations, for example, we can anticipate a situation
where the voters are in an environment that encourages certain type
of interaction (homophily or heterophily). This flexibility in interpretation
renders the model relevant in the context of cultural diversity.

We study the system by numerical analysis and focus on the long-time
behavior of the system, and the distribution of the time to consensus
of opinion. We like to know if there exist stable states and if so,
their nature and their distribution of opinions. We also like to know
the various features of the long time behavior of the system as a
function of the parameters, $p,q,N$, in our model.
Here the \emph{consensus state} is defined to be when all agents in the population are holding the same opinion.
The consensus state is an absorbing state.
 Therefore,
in the simulation, when the population reaches the consensus state,
the simulation ends because from then on the population would not
change. 
Although every simulation will reach the consensus state, the time it takes could be very long. 
The time it takes for the population to evolve to a consensus state (there is only one opinion in the population), from an initial conditions where different opinions are uniformly distributed across the population (or other different initial conditions, depending on the context), is defined as the \emph{time to consensus}. 
Time to consensus is a random variable, and its distribution depends on the particular opinion formation model its parameters. 
The distribution of time to consensus could have significant implications in the
behaviors of the system being modeled. For real election, which has
a deadline for voting, the convergence time is of great practical
importance, as they will determine which party will win the election.

\section{Simulation Results\label{sec:Simulation}}

The Monte Carlo simulation adopts a random sequential updating scheme.
For a given number of available opinions $c$, the \textit{uniform initial condition}
is defined as $N/c$ agents holding opinion $\sigma=1,\cdots,c$, which will be the initial condition  used throughout this work. 
In one Monte Carlo step, an agent $i$ is randomly selected. We consider
all pairs, ($i,j$), with $j\neq i$, and decide whether to establish
link between each pair, according to the following rules: if $\sigma_{i}=\sigma_{j}$,
the two nodes are linked with probability $p$; if $\sigma_{i}\neq\sigma_{j}$,
they are linked with probability $q$. Once all choices are made with the temporary links, $\sigma_{i}$ is updated following a plurality rule: if there exists a plurality opinion $\alpha^{*}$ such that
\begin{equation}
N_{\alpha^{*}}>N_{\beta}\quad\forall\beta\neq\alpha^{*},
\end{equation}
where we assign $\sigma_{i}=\alpha^{*}$. Here $N_{\alpha}$ is the number
of opinion $\alpha$ in the neighborhood. 
Otherwise we will not update the opinion of agent $i$. 
The temporary linking information will be discarded after their updating procedure before the next Monte Carlo step. 
In one Monte Carlo sweep, we perform $N$ Monte Carlo steps. 
In our analysis, the unit of time is one Monte Carlo sweep which corresponds to one MC step per site on average. 

The \textit{time to consensus} is the time it takes for population to evolve from the uniform initial condition to the consensus state.
The distribution of the time to consensus depends on both the number of available opinions $c$ and their initial distribution.
In this work, unless stated otherwise, we always assume the \textit{uniform initial condition} defined above.
We use $T_{c\rightarrow1}$ to denote the time to consensus for $c$-state model. 
To emphasize the uniform initial condition, we may write $T_{c\rightarrow1|N_1=N_2=\cdots=N_c}$, but since we mainly concerns about uniform initial condition, when we write $T_{c\rightarrow1}$, the uniform initial condition is assumed.
For example, the time to consensus for a two-state (three state) model is denoted by $T_{2\rightarrow1}$ ($T_{3\rightarrow1}$).
For two-state model, because of the symmetry of the system, $T_{2\rightarrow 1|N_1=m,N_2=N-m}=T_{2\rightarrow 1|N_1=N-m,N_2=m}$, in the sense that the two random variables have the same probability distribution.
Therefore, we will just write $T_{2\rightarrow 1|m}\equiv T_{2\rightarrow 1|N_1=m,N_2=N-m}=T_{2\rightarrow 1|N_1=N-m,N_2=m}$ to make the notation simpler.
Since time to consensus is a random variable, we use the empirical cumulative distribution function (ECDF) to visualize
the distribution. ECDF is defined as $F_c(t)\equiv\textrm{Prob}(T_{c\rightarrow1}\le t)$. In Fig.~\ref{fig:ECDF_p_0.35_0.5_0.65}
we show the time to consensus for three different values of $p$. Small
or large values of $p$ result in longer time to consensus. 
We summarize the result in Table~\ref{tab:avg_T_state_two_three} for the average time to consensus $\left\langle T_{3\rightarrow1}\right\rangle $ and $\left\langle T_{2\rightarrow1}\right\rangle $.
The ECDF of time to consensus for two-state model is also shown. When
$p$ is around 0.5, times to consensus for two-state and three-state
population are similar in distribution. However, when $p$ is small,
e.g., $0.35$, time to consensus for three-state population is statistically
shorter than that of a two-state population in the sense that at any
time point $t_{1}$, the probability that a three-state population
has reached the consensus state is larger than that of a two-state
system. When $p$ is large, e.g., 0.65, the situation is similar. The
shortening of time to consensus when $p=0.35$ or $p=0.65$ is even
more prominent when the number of available opinions is larger. (Figs.~\ref{fig:ECDF_p_0.35_c_2_3_5_6_10}
and \ref{fig:ECDF_p_0.65_c_2_3_5_6_10}) These results may go against
intuitions.

\begin{center}
\begin{figure}
\begin{centering}
\includegraphics[width=1\columnwidth]{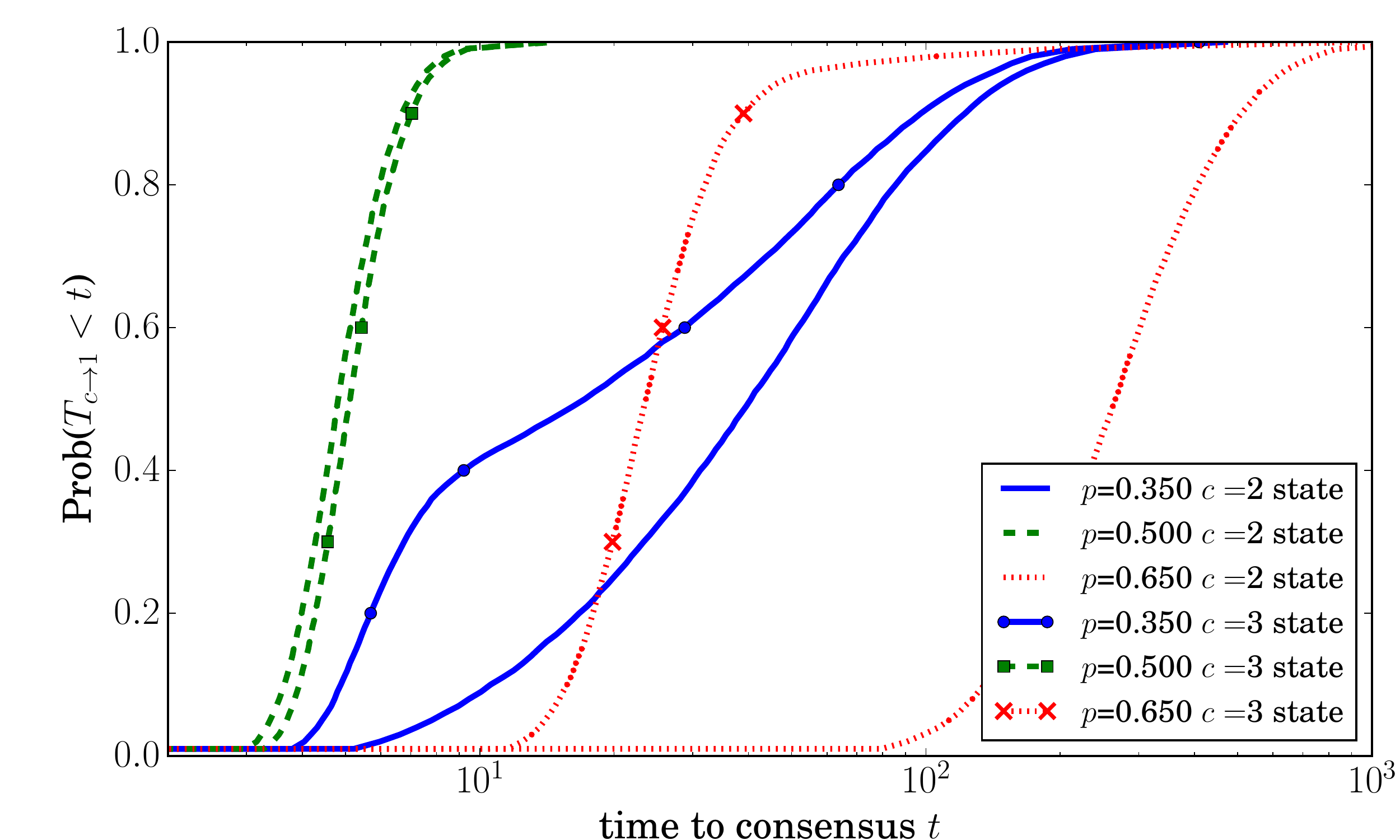}
\par\end{centering}

\protect\caption{\label{fig:ECDF_p_0.35_0.5_0.65}Empirical cumulative distribution
function of time to consensus. $N=150$.}

\end{figure}

\par\end{center}

\begin{center}
\begin{table}
\begin{centering}
\begin{tabular}{|c|c|c|}
\hline 
 & two-state & three-state\tabularnewline
\hline 
\hline 
$p=0.35$ & 56.5 & 38.4\tabularnewline
\hline 
$p=0.5$ & 5.0 & 5.3\tabularnewline
\hline 
$p=0.65$ & 299.6 & 30.4\tabularnewline
\hline 
\end{tabular}
\par\end{centering}

\protect\caption{\label{tab:avg_T_state_two_three}Average time to consensus $\left< T_{c\rightarrow1}\right>$
for various values of $p$ for two-state and three-state models. $N$=150.}

\end{table}

\par\end{center}

\begin{center}
\begin{figure}
\begin{centering}
\includegraphics[width=0.8\columnwidth]{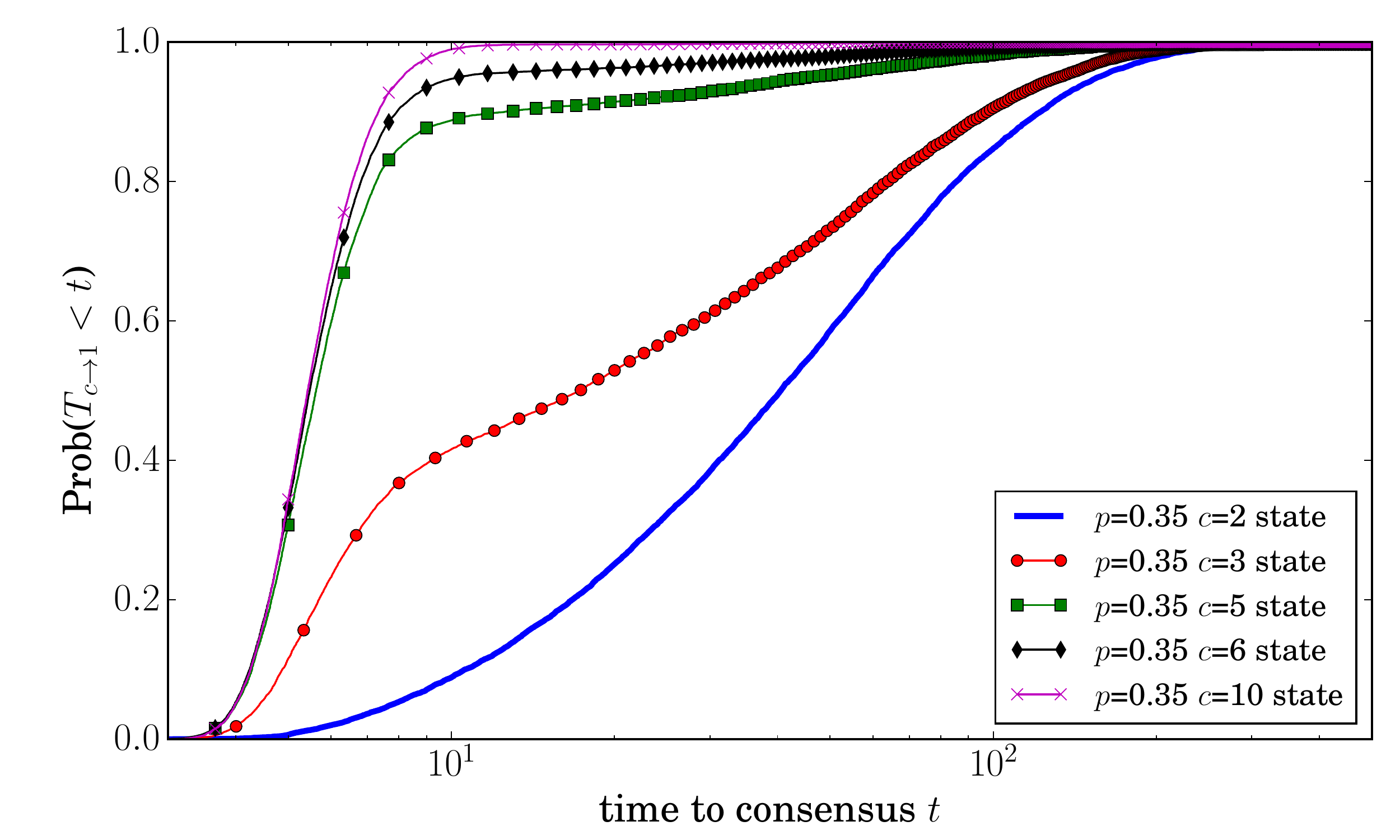}
\par\end{centering}

\protect\caption{\label{fig:ECDF_p_0.35_c_2_3_5_6_10}Empirical cumulative distribution
function of time to consensus. $N=150,p=0.35$.}

\end{figure}

\par\end{center}

\begin{center}
\begin{figure}
\begin{centering}
\includegraphics[width=0.8\columnwidth]{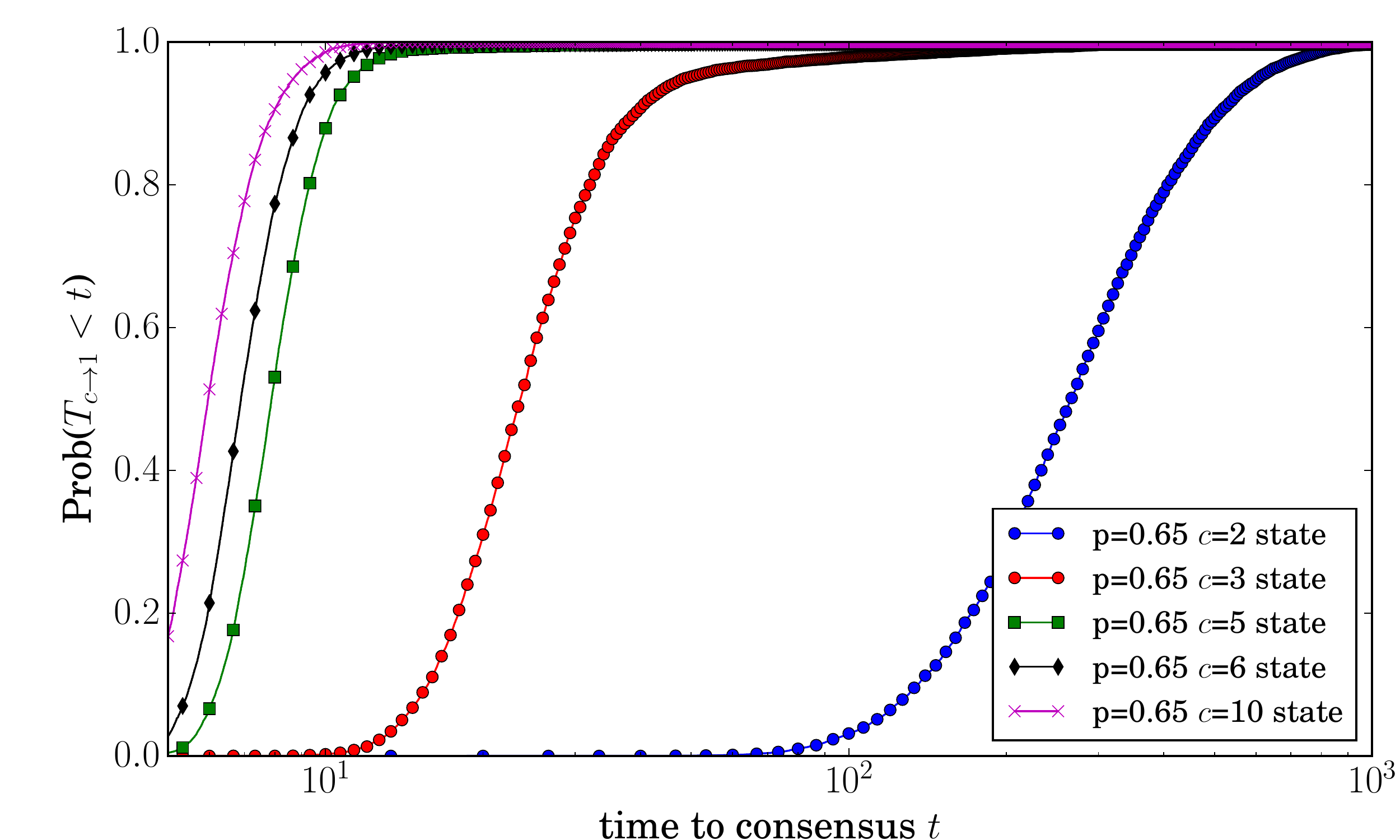}
\par\end{centering}

\protect\caption{\label{fig:ECDF_p_0.65_c_2_3_5_6_10}Empirical cumulative distribution
function of time to consensus. $N=150,p=0.65$.}
\end{figure}

\par\end{center}

\begin{center}
\begin{table}
\begin{centering}
\begin{tabular}{|c|c|c|c|c|c|}
\hline 
 & two-state & three & five & six & ten\tabularnewline
\hline 
\hline 
$p=0.35$ & 56.5 & 38.4 & 11.5 & 8.2 & 5.8\tabularnewline
\hline 
$p=0.65$ & 299.6 & 30.4 & 8.7 & 7.4 & 6.2\tabularnewline
\hline 
\end{tabular}
\par\end{centering}

\protect\caption{\label{tab:avg_T_state_235610}Average time to consensus $\left< T_{c\rightarrow1}\right>$
of two-, three-, five-, six- and ten-state models. $N=150$.}

\end{table}

\par\end{center}

To understand the mechanism of the acceleration, it is helpful to break down the whole process into two subprocesses: 
\begin{itemize}
\item Process \rom{1}: one of the three opinions goes extinct. 
\item Process \rom{2}: the population with the two remaining opinions finally reaches the consensus state. 
\end{itemize}
The time process \rom{1} takes will be referred to as the third-opinion extinction time $T_{3\rightarrow2,m}$, where $m$ indicates the distribution of the two remaining opinions at the end of the process. 
Similarly, the time of process \rom{2} takes will be denoted by $T_{2\rightarrow1|m}$, where $m$ indicates the distribution of the two remaining opinions at the beginning of the process. 

The conditional probability distribution of the time to consensus of a two state population given an non-uniform initial condition is denoted by $P_{2\rightarrow1|m}(T)$, and the distribution of the third-opinion extinction time given the final condition $m$ will be referred to as $P_{3\rightarrow2,m}(T)$.
Therefore, we write $T_{3\rightarrow1}\equiv T_{3\rightarrow2,m}+T_{2\rightarrow1|m}$, in the sense that
\begin{equation}
P_{3\rightarrow1}(T)=\sum_{t=0}^T\sum_{a=0}^N P_{3\rightarrow2,a}(T-t)P_{2\rightarrow1|a}(t)P_m(a),
\end{equation}
where $P_m(a)$ is the probability that $m=a$ at the end of process \rom{1}.
We define $\left<T_{3\rightarrow2}\right>=\left<T_{3\rightarrow2,m}\right>_m$. 
Fig.~\ref{fig:avg_T_3to2} shows that, if $p<0.5$, $\left<T_{3\rightarrow2}\right>$ is insensitive to $N$ and $p$, but if $p>0.5$, $\left<T_{3\rightarrow2}\right>$ scales as $\exp(N)$. 

\begin{center}
\begin{figure}
\begin{centering}
\includegraphics[width=0.8\columnwidth]{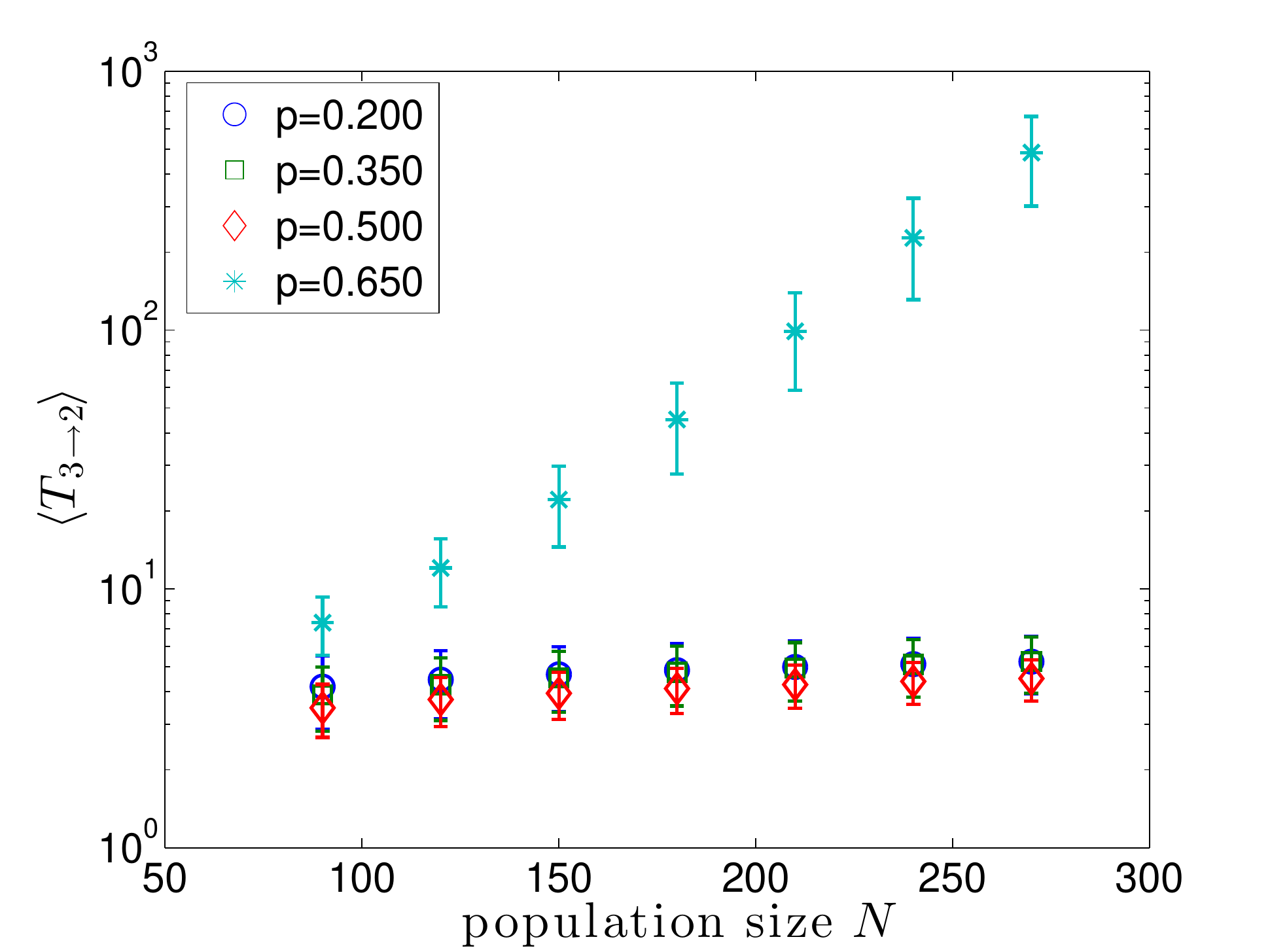}
\par\end{centering}

\protect\caption{\label{fig:avg_T_3to2}$\left<T_{3\rightarrow2}\right>$ as a function
of $N$ for various values of $p$. The error bars are the standard
deviations of $\left<T_{3\rightarrow2}\right>$. $\left<T_{3\rightarrow2}(p=0.65)\right>\sim\exp(0.026N)$.}

\end{figure}

\par\end{center}

\section{Master Equation\label{sec:M-Equation}}

To have a better understanding of the dynamics, we investigate the master equation of the three-opinion system. 
Denote the number of agents holding opinion 1 by $N_1$, opinion 2 by $N_2$ and opinion 3 by $N_3$.
The configuration of the population can be described by these three numbers $(N_1,N_2,N_3)$. 
Since $N_1+N_2+N_3=N$, where $N$ is the size of the population, two numbers out of three suffice. 

The master equation of the system is of the following form
\begin{align}
\begin{split}
 &\partial_{t}P(N_1,N_2)\\
 =&-\dots P(N_1,N_2)\\
 & +\cdots P(N_1+1,N_2-1)+\cdots P(N_1+1,N_2)\\
 & +\cdots P(N_1-1,N_2+1)+\cdots P(N_1,N_2+1)\\
 & +\cdots P(N_1-1,N_2)+\cdots P(N_1,N_2-1),
\end{split}
\end{align}
where the first term on the RHS is the outflow and other terms are the inflows. 
Here the $\cdots$ are the coefficients involving transition probabilities.
We now introduce the transition probability $W_{1\rightarrow 2}(N_1,N_2)$ for a particular opinion 1 to change into opinion 2 after $\Delta t$.
According to the update rule of the model, $W_{1\rightarrow2}(N_1,N_2)$ is the product of the probability that an agent holding opinion 1 is chosen in the current round, and the probability that in the temporary neighborhood, the number of agents holding opinion 2 ($l'$), is larger than the number of agents with opinion 1 ($l$), as well as the number of agents holding opinion 3 ($l''$). 
One can find that theses probabilities can be written in the form of Binomial distribution:
\begin{equation}
B_{n,p}(l)={n \choose l}p^{l}(1-p)^{n-l},
\end{equation}
so that the product of the probabilities yield the following expression for $W_{1\rightarrow2}(N_1,N_2)$
\begin{align}
\begin{split}
&W_{1\rightarrow2}(N_1,N_2)\\
=& \frac{N_1}{N}\sum_{l=0}^{N_1-1}\sum_{l'=0}^{N_2}\sum_{l''=0}^{N_3}B_{N_1-1,p}(l)B_{N_2,q}(l')B_{N_3,q}(l'')\\
 & \Theta(l'-l)\Theta(l'-l'').
\end{split}
\end{align}

Now $W_{1\rightarrow3}(N_1,N_2)$,~$W_{2\rightarrow1}(N_1,N_2)$, etc., can be derived in similar fashion. 

The complete M-equation is therefore

\begin{widetext}

\begin{align}
\begin{split}
&\partial_{t}P(N_1,N_2)\\
= &-\left[W_{1\rightarrow2}(N_1,N_2)+W_{1\rightarrow3}(N_1,N_2)+W_{2\rightarrow1}(N_1,N_2)\right.\\
&\left.+W_{2\rightarrow3}(N_1,N_2)+W_{3\rightarrow1}(N_1,N_2)+W_{3\rightarrow2}(N_1,N_2)\right]P(N_1,N_2)\\
 & +W_{1\rightarrow2}(N_1+1,N_2-1)P(N_1+1,N_2-1)+W_{1\rightarrow3}(N_1+1,N_2)P(N_1+1,N_2)\\
 & +W_{2\rightarrow1}(N_1-1,N_2+1)P(N_1-1,N_2+1)+W_{2\rightarrow3}(N_1,N_2+1)P(N_1,N_2+1)\\
 &+W_{3\rightarrow1}(N_1-1,N_2)P(N_1-1,N_2)+W_{3\rightarrow2}(N_1,N_2-1)P(N_1,N_2-1).
\end{split}
\end{align}
\end{widetext}
Note that $P(N_1,N_2)$ is time-dependent unless otherwise stated.

Because of the symmetry in the transition probability, $W_{1\rightarrow3}(N_1,N_2)$, $W_{2\rightarrow1}(N_1,N_2)$ can be obtained by simple transformation of $W_{1\rightarrow2}(N_1,N_2)$. 
Please refer to the master equation that only contains $W_{1\rightarrow2}$ in the supplementary materials.

To appreciate the qualitative feature of $W_{1\rightarrow2}$ when $N$ is finite, we show the ternary contour plot in Fig.~\ref{fig:W12_N_900_p_0.350}.
Here we introduce a change of variable $x=N_1/N,y=N_2/N,z=N_3/N$ so that we can investigate population-size independent phenomena more clearly.
The dependence of $W_{1\rightarrow2}$ on $x$ is evident. Significant variations
in magnitude of $W_{1\rightarrow2}$ concentrates in a region, outside of which
$W_{1\rightarrow2}$ is very close to zero. Therefore, for qualitative analysis,
the region where $W_{1\rightarrow2}$ is effectively zero will be referred to
as the zero region, and the remaining region is called the positive
region (see Fig.~\ref{fig:W12_p_0.35_positive}). 
In fact, we show in the supplementary materials that in the large $N$ limit, the boundary between the positive region and the zero region is $y=\max[(x-\epsilon)p/q, (1-x)/2]$, where $\epsilon=1/N$.
Inside the positive region, $W_{1\rightarrow2}(x,y)=x$ and vanishes in the zero region. 
Changing $p$ changes the boundary between the two regions.

\begin{center}
\begin{figure}
\begin{centering}
\includegraphics[width=1\columnwidth]{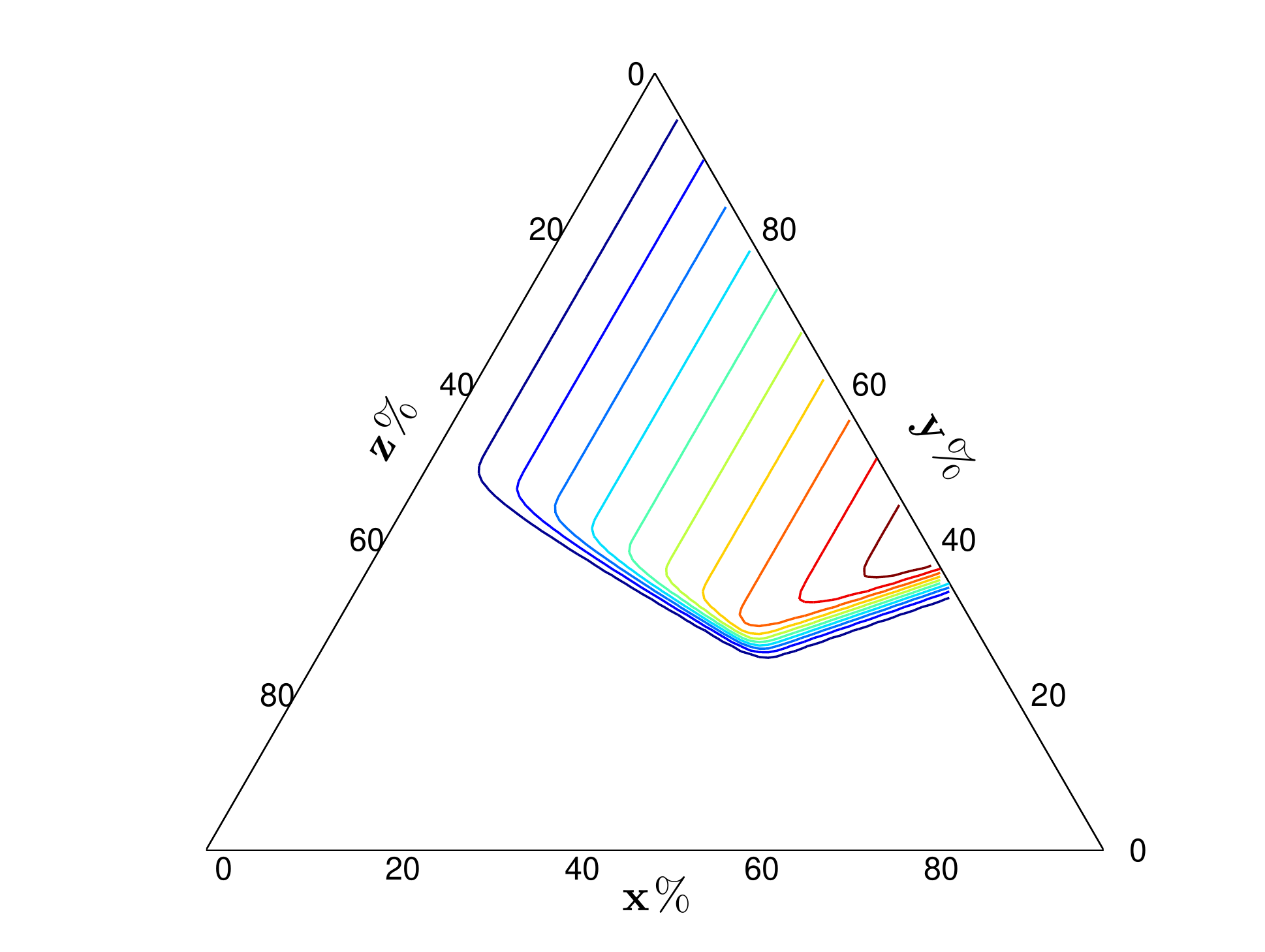}
\par\end{centering}

\protect\caption{\label{fig:W12_N_900_p_0.350}$W_{1\rightarrow2}$ on a ternary contour plot.
The ticks in the three axes are percentages. $N=900,p=0.35$.}

\end{figure}

\par\end{center}

\begin{center}
\begin{figure}
\begin{centering}
\includegraphics[width=0.9\columnwidth]{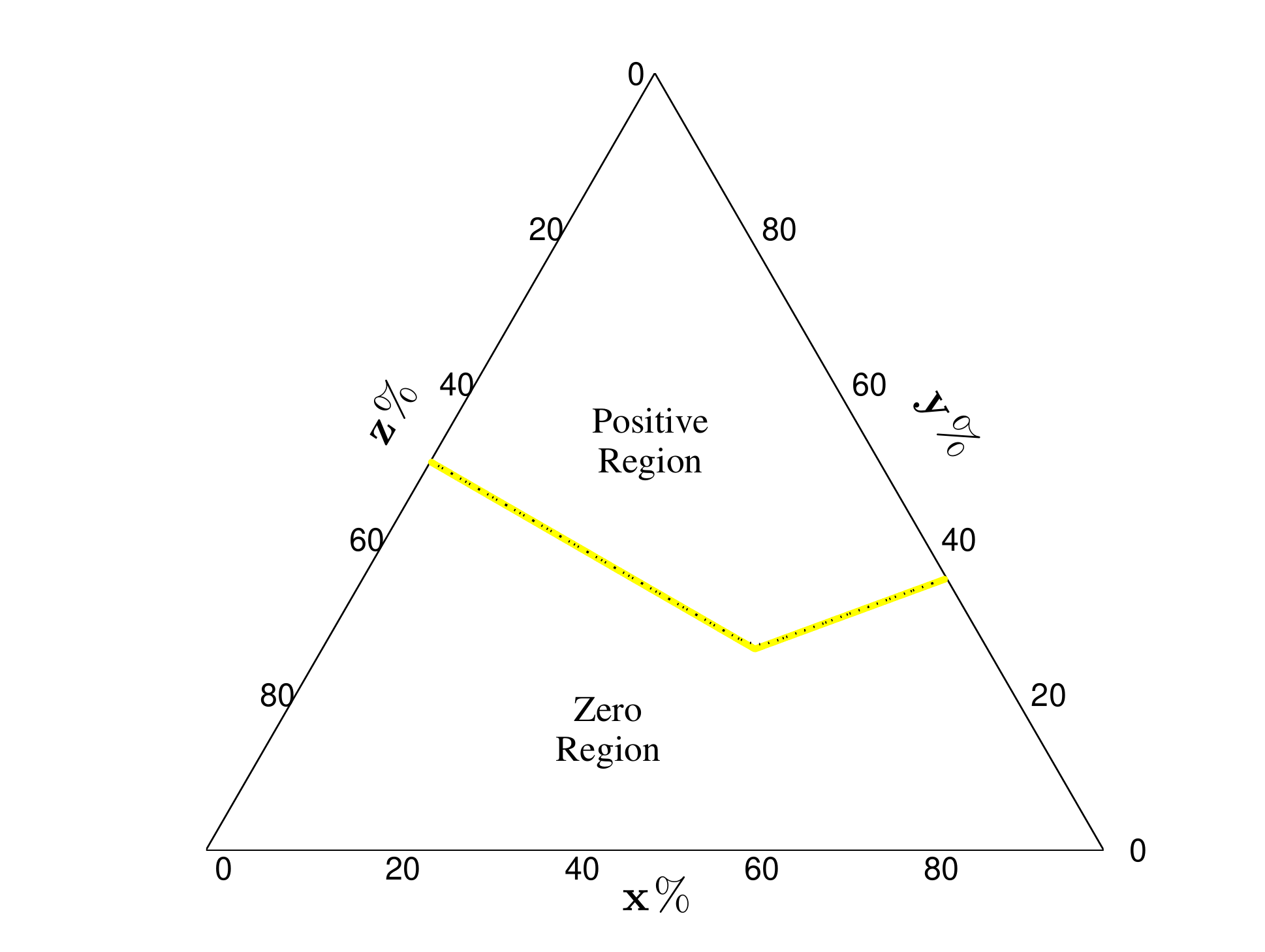}
\par\end{centering}

\protect\caption{\label{fig:W12_p_0.35_positive}As $N\rightarrow\infty$, $W_{1\rightarrow2}(N_1,N_2)$
can be divided into two regions: a region where $W_{1\rightarrow2}$ is significant
larger than zero (positive region), and one where $W_{1\rightarrow2}$ is effectively
zero. $N=900,p=0.35$. Black dotted line is the boundary calculated from numerical data and the yellow solid line is from the theoretical calculation (see the supplementary materials for the derivation).}

\end{figure}

\par\end{center}

Next we introduce the definition of the region in $(x,y)$ where the transition probability is positive.
Suppose $\Omega=\{(x,y)|0\le x\le 1,0\le y\le 1,x+y\le 1\}$, $\Gamma=\{(1,2),(1,3),(2,1),(2,3),(3,1),(3,2)\}$,
$\mathbb{F}\left[(i,j)\right]=\{(x,y)|(x,y)\in\Omega,W_{ij}(x,y)>0\}$
and $\gamma$ is a member of the powerset of $\Gamma$, $2^{\Gamma}$,
we define
\begin{equation}
\mathbb{A}(\gamma)=\bigcap_{(i,j)\in\gamma}\mathbb{F}\left[(i,j)\right]\bigcap_{(i,j)\in\Gamma\setminus\gamma}\mathbb{F}^{C}[(i,j)],
\end{equation}
where $\mathbb{F}^{C}[(i,j)]\equiv\Omega\setminus\mathbb{F}\left[(i,j)\right]$, which is the area where $W_{i\rightarrow j}=0$.
In finite-size system, however, $W_{i\rightarrow j}(x,y)$ is never exactly zero
except at some points on the boundary of $\Omega$. A threshold $\Delta$ needs to be chosen such that whenever $W_{1\rightarrow 2}(N_1,N_2)<\Delta$, we assume $W_{1\rightarrow 2}(N_1,N_2)=0$.

One can show that the family of sets $\left\{ \mathbb{A}(\gamma)\right\} _{\gamma\in\Gamma}$
is a partition of the set $\Omega$, and a member in the partition
is called a block. Fig.~\ref{fig:trans_prob_regions_p_0.35} shows
the partition when $p=0.35$. There are at least 9 qualitatively distinct
regions.

\begin{center}
\begin{figure}
\begin{centering}
\includegraphics[width=0.95\columnwidth]{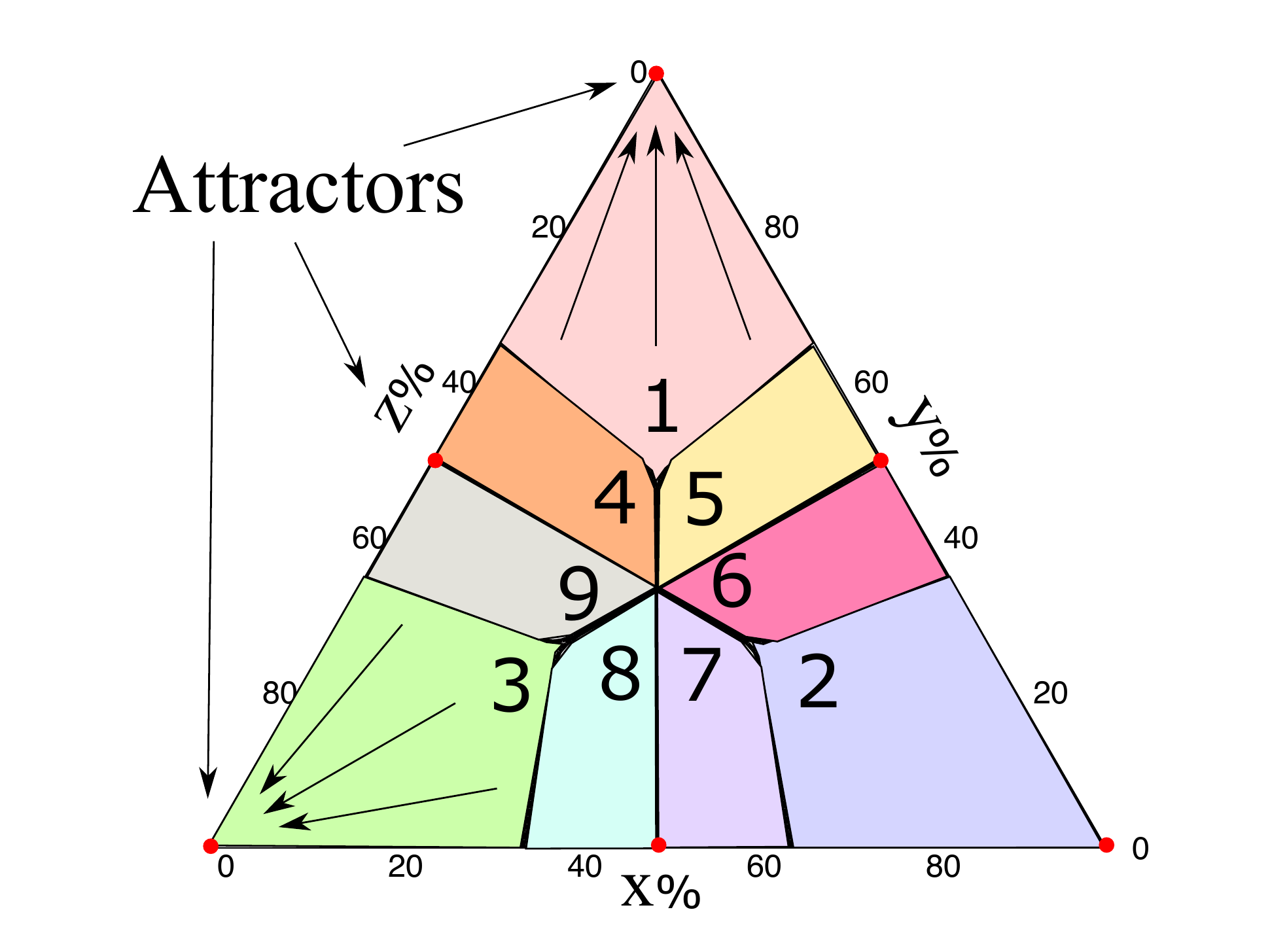}
\par\end{centering}

\protect\caption{\label{fig:trans_prob_regions_p_0.35}$W_{1\rightarrow2}(N_1,N_2)$ on a ternary
plot with partition $\left\{ \mathbb{A}(\gamma)\right\} $.$N=900,p=0.35$.
Region 1 is $\mathbb{A}\left[(1,2),(3,2)\right]$. Region 2 is $\mathbb{A}\left[(2,1),(3,1)\right]$.
Region 3 is $\mathbb{A}\left[(2,3),(1,3)\right]$. Region 4 is $\mathbb{A}\left[(3,2),(1,2)\right]$.
Region 5 is $\mathbb{A}\left[(2,1),(3,2),(1,2)\right]$. Region 6
is $\mathbb{A}\left[(3,1),(2,1),(1,2)\right]$. Region 7 is $\mathbb{A}\left[(1,3),(2,1),(3,1)\right]$.
Region 8 is $\mathbb{A}\left[(3,1),(1,3),(2,3)\right]$. Region 9
is $\mathbb{A}\left[(3,2),(1,3),(2,3)\right]$. Numerical results
suggest that the unnumbered regions/sets will vanish in relative area
as $N\rightarrow\infty$. The location of attractors in $N\rightarrow\infty$
are marked.}
\end{figure}

\par\end{center}

In region 1 of Fig.~\ref{fig:trans_prob_regions_p_0.35}, $\mathbb{A}\left[(1,2),(3,2)\right]$, $W_{1\rightarrow2}$ and $W_{3\rightarrow2}$ are positive and proportional to $x$ and $z$, respectively, and the master equation can therefore be approximated by (refer to supplementary material):
\begin{equation}
\partial_{t}P(x,y)\approx \epsilon\left[x\frac{\partial P}{\partial x}+(y-1)\frac{\partial P}{\partial y}+2P(x,y)\right]
\end{equation}
where $\epsilon=1/N$ and the last approximation keeps
only the first order terms of $\epsilon$. Therefore, in the limit $N\rightarrow\infty$,
or equivalently $\epsilon\rightarrow0$, the diffusion of $P(x,y)$
can be ignored. Using the method of characteristics, it can be shown
that the solution $P(x,y)\sim f((y-1)/x,t+\log x)$,
where $f(\cdots,\cdots)$ is an arbitrary two-variable function so
that if $P(x,y,t=0)=\delta(x-x_{0},y-y_{0})$,
the probability mass will travel on the trajectory $x(t)=x_{0}\exp(-t),y(t)=(y_{0}-1)\exp(-t)+1$,
or in other words, $P(x,y)$ that starts at any point
inside region 1 goes to $(x=0,y=1)$ exponentially fast
and $(x=0,y=1)$ is the attractor of region 1. 

Similarly, in region 4 of Fig.~\ref{fig:trans_prob_regions_p_0.35}, or $\mathbb{A}\left[(3,2),(1,2)\right]$,
only $W_{1\rightarrow2},W_{2\rightarrow3}$ and $W_{3\rightarrow2}$ are positive, and the master equation
in that region can be approximated by, in the $N\rightarrow\infty$
limit (refer to supplementary material), 
\begin{equation}
\partial_{t}P(x,y) \approx \epsilon\left[2P(x,y)+x\frac{\partial P}{\partial x}+(2y-1)\frac{\partial P}{\partial y}\right].
\end{equation}

The solution is $P(x,y)\sim f((2y-1)/2x^{2},t+\log x)$,
so that $x(t)=x_{0}\exp(-t),y(t)=1/2+(y_{0}-1/2)\exp(-2t)$
and $(x=0,y=1/2)$ is the attractor of region 4. The results above
show that the time scales in which the population move through regions
1 and 4, are independent of the population size $N$, in agreement
with the results in Fig.~\ref{fig:avg_T_3to2} for those cases with $p$ less than 0.5.

Suppose $(x_{0},y_{0})$ is inside one of the numbered region, the
time evolution $P(N_1,N_2)$ with $P(x,y,t=0)=\delta(x-x_{0})\delta(y-y_{0})$
has two features: 1) original delta function-like $P(N_1,N_2)$ distribution
will spread out 2) there is an overall motion of the distribution.
The arrow in the region denotes the overall direction of motion of
$P(N_1,N_2)$ in that region.
To see more clearly the evolution of the probability distribution, we combine the analytical results of the master equation with numerical simulation. 
When $p$ is small, after an initial diffusion of the probability
distribution, the probability mass will be split into 3 parts, which
will then pass through regions 4,5,6,7,8,9. ($p=0.29$ in Fig.~\ref{fig:time_evo_N_150_p_0.29})
As $p$ increases, a new pattern emerges: the probability mass will
be broken into six pieces, and significant amounts of probability
mass will pass through region 1,2 and 3 ($p=0.35$ in Fig.~\ref{fig:time_evo_N_150_p_0.35}).

\begin{center}
\begin{figure}
\begin{centering}
\subfloat[$t=0.2$.]{\begin{centering}
\includegraphics[width=0.3\columnwidth]{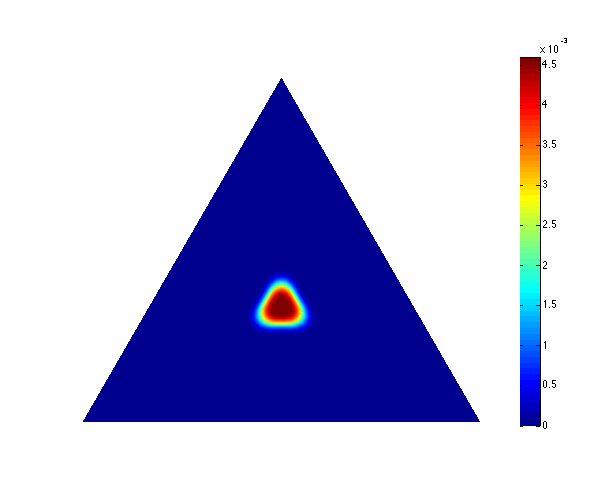}
\par\end{centering}

}\subfloat[$t=0.4$.]{\begin{centering}
\includegraphics[width=0.3\columnwidth]{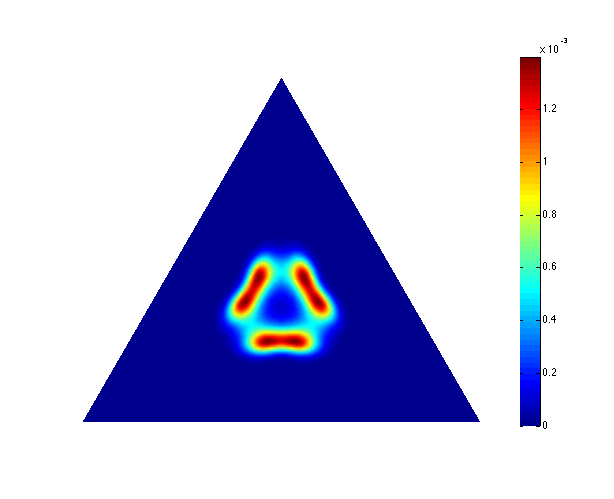}
\par\end{centering}

}\subfloat[$t=1$.]{\begin{centering}
\includegraphics[width=0.3\columnwidth]{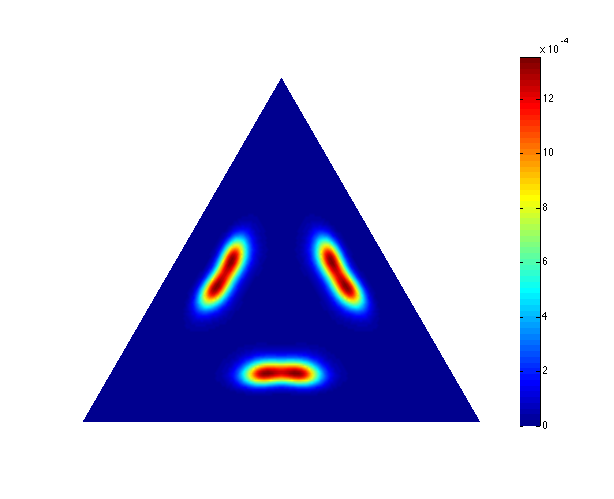}
\par\end{centering}

}
\par\end{centering}

\protect\caption{\label{fig:time_evo_N_150_p_0.29}$N=150,p=0.29$.}

\end{figure}

\par\end{center}

\begin{center}
\begin{figure}
\begin{centering}
\subfloat[$t=0.2$.]{\begin{centering}
\includegraphics[width=0.3\columnwidth]{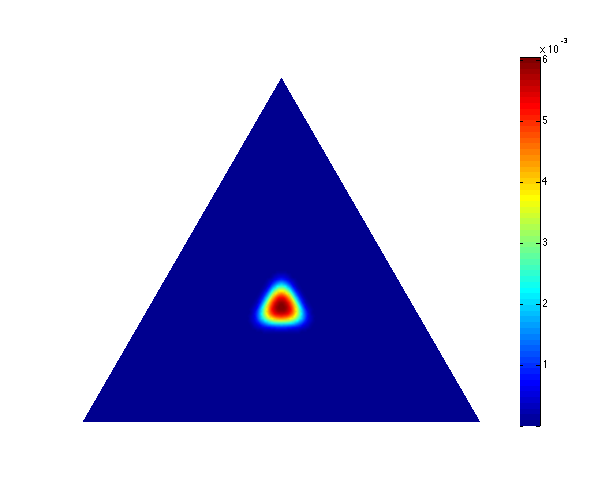}
\par\end{centering}

}\subfloat[$t=0.4$.]{\begin{centering}
\includegraphics[width=0.3\columnwidth]{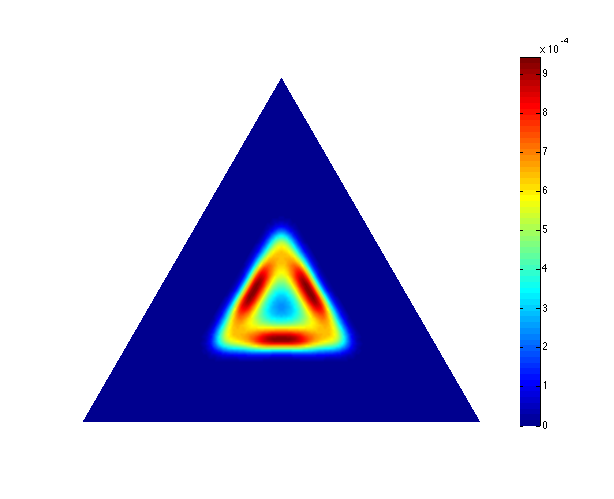}
\par\end{centering}

}\subfloat[$t=13/15$.]{\begin{centering}
\includegraphics[width=0.3\columnwidth]{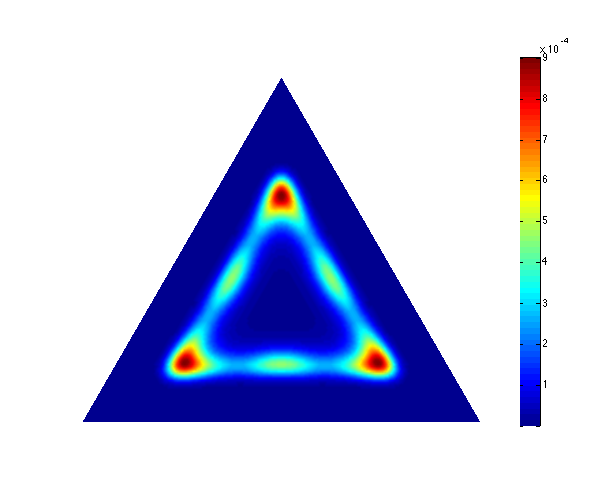}
\par\end{centering}

}
\par\end{centering}

\protect\caption{\label{fig:time_evo_N_150_p_0.35}$N=150,p=0.35.$}
\end{figure}

\par\end{center}

As $p$ increases to 0.5, regions 1, 2 and 3 become larger, while
regions 4,5,6,7,8 and 9 become smaller accordingly. When $p=0.5$,
regions 4,5,6,7,8 and 9 become so small that they do not have significant
effects on the dynamics. See Fig. \ref{fig:trans_prob_regions_p_0.5}.
As a result, major parts of the probability mass will go through regions
1,2,3 only.

\begin{center}
\begin{figure}
\begin{centering}
\subfloat[$t=0.2$.]{\begin{centering}
\includegraphics[width=0.3\columnwidth]{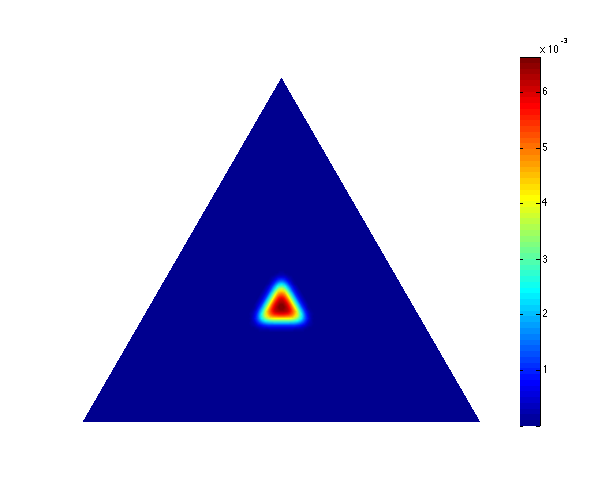}
\par\end{centering}

}\subfloat[$t=0.4$.]{\begin{centering}
\includegraphics[width=0.3\columnwidth]{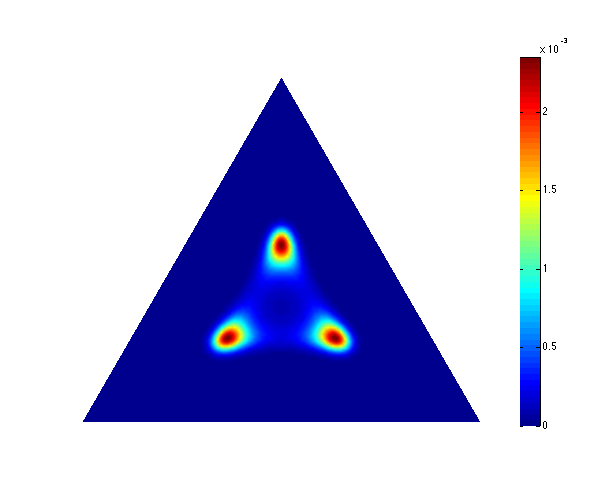}
\par\end{centering}

}\subfloat[$t=13/15$.]{\begin{centering}
\includegraphics[width=0.3\columnwidth]{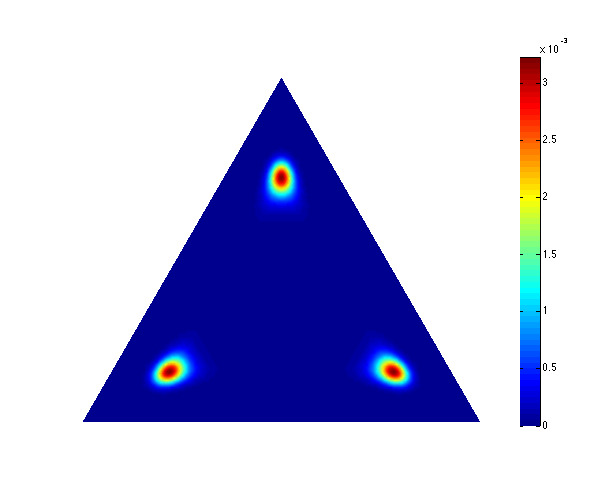}
\par\end{centering}

}
\par\end{centering}

\protect\caption{\label{fig:time_evo_N_150_p_0.5}$N=150,p=0.5$.}
\end{figure}

\par\end{center}

\begin{center}
\begin{figure}
\begin{centering}
\includegraphics[width=0.95\columnwidth]{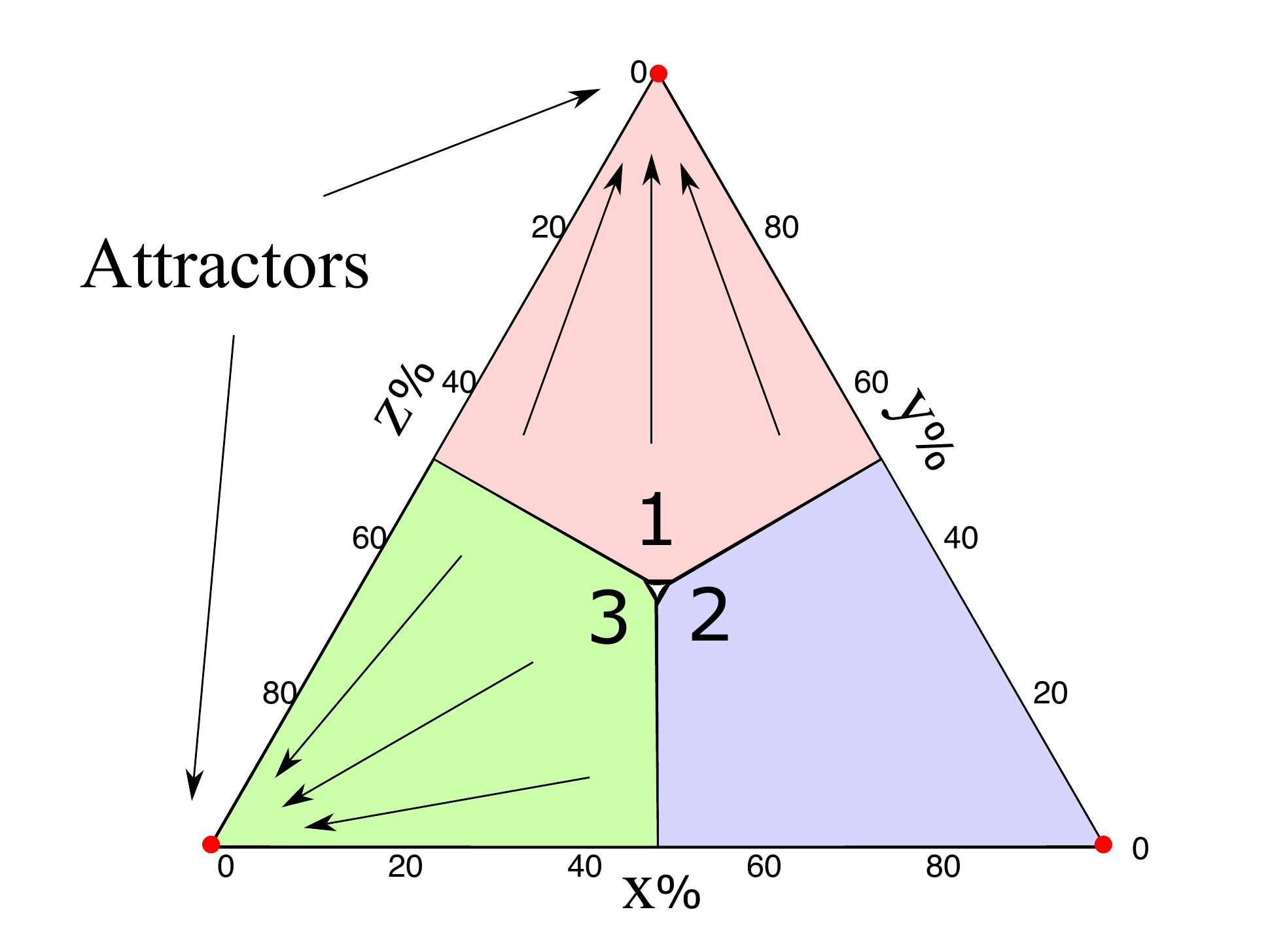}
\par\end{centering}

\protect\caption{\label{fig:trans_prob_regions_p_0.5}$W_{1\rightarrow2}(N_1,N_2)$ on a ternary plot
with partition $\left\{ \mathbb{A}(\gamma)\right\} $. $N=900,p=0.5$.
The numbering scheme is the same as in Fig.~\ref{fig:W12_N_900_p_0.350}.
The location of attractors in $N\rightarrow\infty$ are marked.}
\end{figure}

\par\end{center}

As $p$ increases beyond 0.5, a new region emerges. See Fig.~\ref{fig:trans_prob_regions_p_0.700}
for region 10, where all transition probabilities are qualitatively
zero. Numerical results show that $P(x,y)$ with uniform initial condition,
i.e., $P(x,y)=\delta(x-1/3)\delta(y-1/3)$, will diffuse and move
to the boundaries of region 10. 

\begin{center}
\begin{figure}
\begin{centering}
\includegraphics[width=0.95\columnwidth]{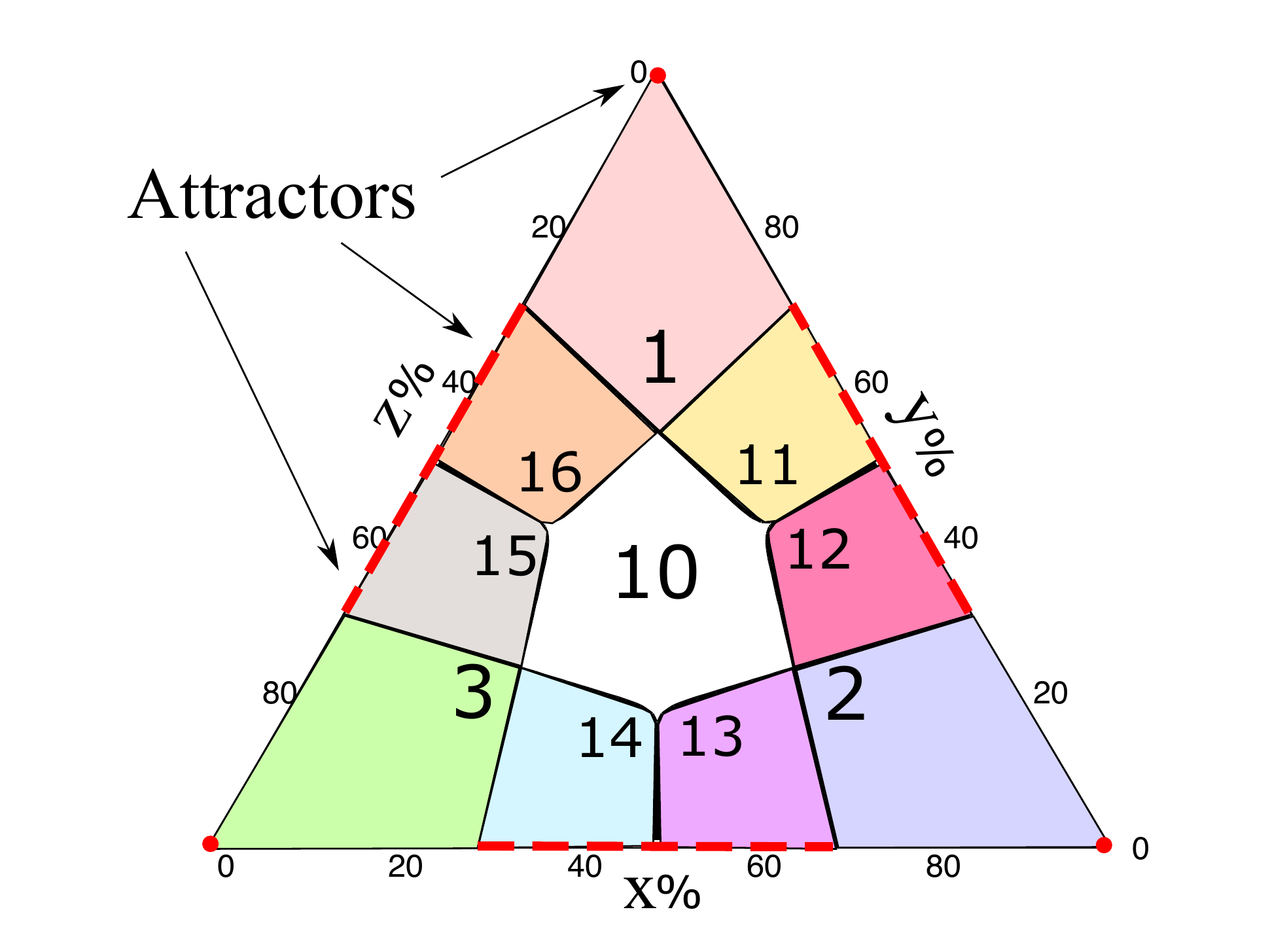}
\par\end{centering}

\protect\caption{\label{fig:trans_prob_regions_p_0.700}$W_{1\rightarrow2}(N_1,N_2)$ on a ternary
plot with partition $\left\{ \mathbb{A}(\gamma)\right\} $. $N=900,p=0.7$.
Region 10 is $\mathbb{A}(\emptyset)$. The location of attractors
in $N\rightarrow\infty$ are marked. The dashed lines mean that each
point on the lines is an attractor.}

\end{figure}

\par\end{center}

\section{Mechanism behind Accelerated Consensus\label{sec:Mechanism}}

After the qualitative analysis of the flow of probability, we now compute the average time to consensus. 
The probability distribution of the time to consensus $P_{3\rightarrow1}(T)$ can
be written as
\begin{align}
P_{3\rightarrow}(T) & =\int_{0}^{T}P_{3\rightarrow2,2\rightarrow1}(T-t,t)dt,
\end{align}
where $P_{3\rightarrow2,2\rightarrow1}(t_{1},t_{2})$ is the joint
probability that process \rom{1} (denoted by $3\rightarrow2$) takes time $t_{1}$ and process
\rom{2} (denoted by $2\rightarrow1$) takes time $t_{2}$. This decomposition can be written as:
\begin{align}
\begin{split}
P_{3\rightarrow1}(T)= & \int_{0}^{N} da \int_{0}^{T}P_{3\rightarrow2,2\rightarrow1,m}(T-t,t,a)dt\\
= & \int_{0}^{N} da \int_{0}^{T}P_{3\rightarrow2,2\rightarrow1}(T-t,t|a)P_{m}(a)dt
\end{split}
\end{align}
where $P_{3\rightarrow2,2\rightarrow1,m}(t_{1},t_{2},a)$ is the joint probability that process \rom{1} takes time $t_{1}$ and process \rom{2} takes time $t_{2}$, and that the number of one of the opinion (because of the symmetry, it does not matter whether $N_1=m$ or $N_2=m$) $m=a$ at the end of process \rom{1}. 
However, knowing $m=a$ decouples process \rom{1} from
process \rom{2}, since the whole process is a Markov process. From
numerical integration of the master equation, we observe results shown
in Fig.~\ref{fig:Pm_p_0.370}. Most of the probability mass either
concentrates near $a=N/2$ or at the two corners.
As $N$ increases, comparatively more probability mass concentrates near the center, and the width of the centering probability mass becomes narrower.
That is to say, as $N\rightarrow\infty$, $P_{m}(a)$ can be approximated as
\begin{equation}
P_{m}(a)\approx\begin{cases}
[1-C(N,p)]/2 & a=0,N\\
C(N,p) & a=N/2\\
0 & \text{otherwise}
\end{cases},\label{eq:P_m_approx}
\end{equation}
where $(1-C(N,p))/2\equiv\int_{0}^{\delta}P_{m}(a)da$, and $1-C(N,p)$ can be
interpreted as the probability that at the end of process \rom{1},
$m=0$ or $N$. This approximation does not hold well when $p>0.5$.

\begin{center}
\begin{figure}
\begin{centering}
\subfloat[\label{fig:Pm_N_300_p_0.370}$N=300$]{\begin{centering}
\includegraphics[width=0.45\columnwidth]{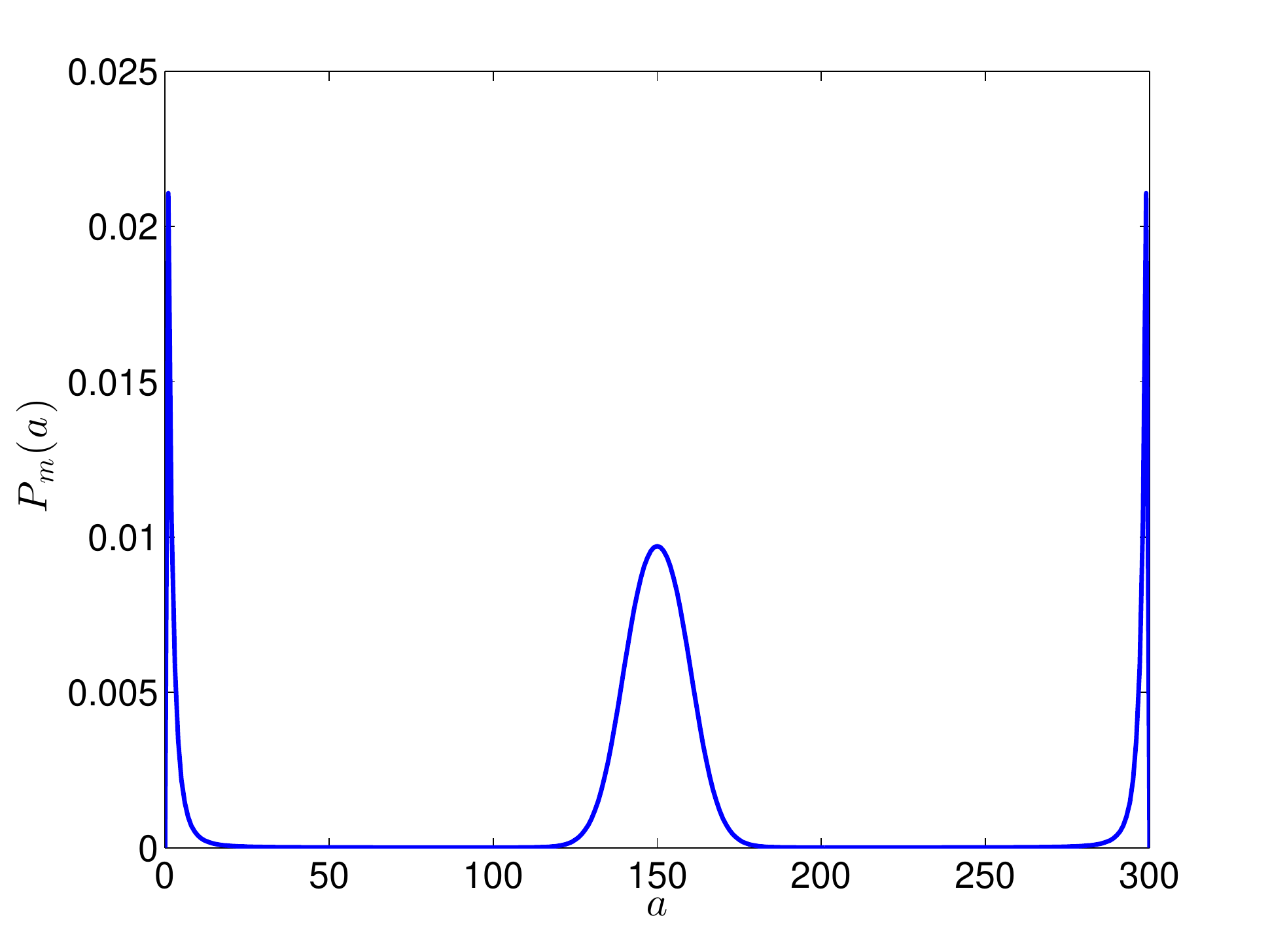}
\par\end{centering}

}\subfloat[\label{fig:Pm_N_900_p_0.370}$N=900$]{\begin{centering}
\includegraphics[width=0.45\columnwidth]{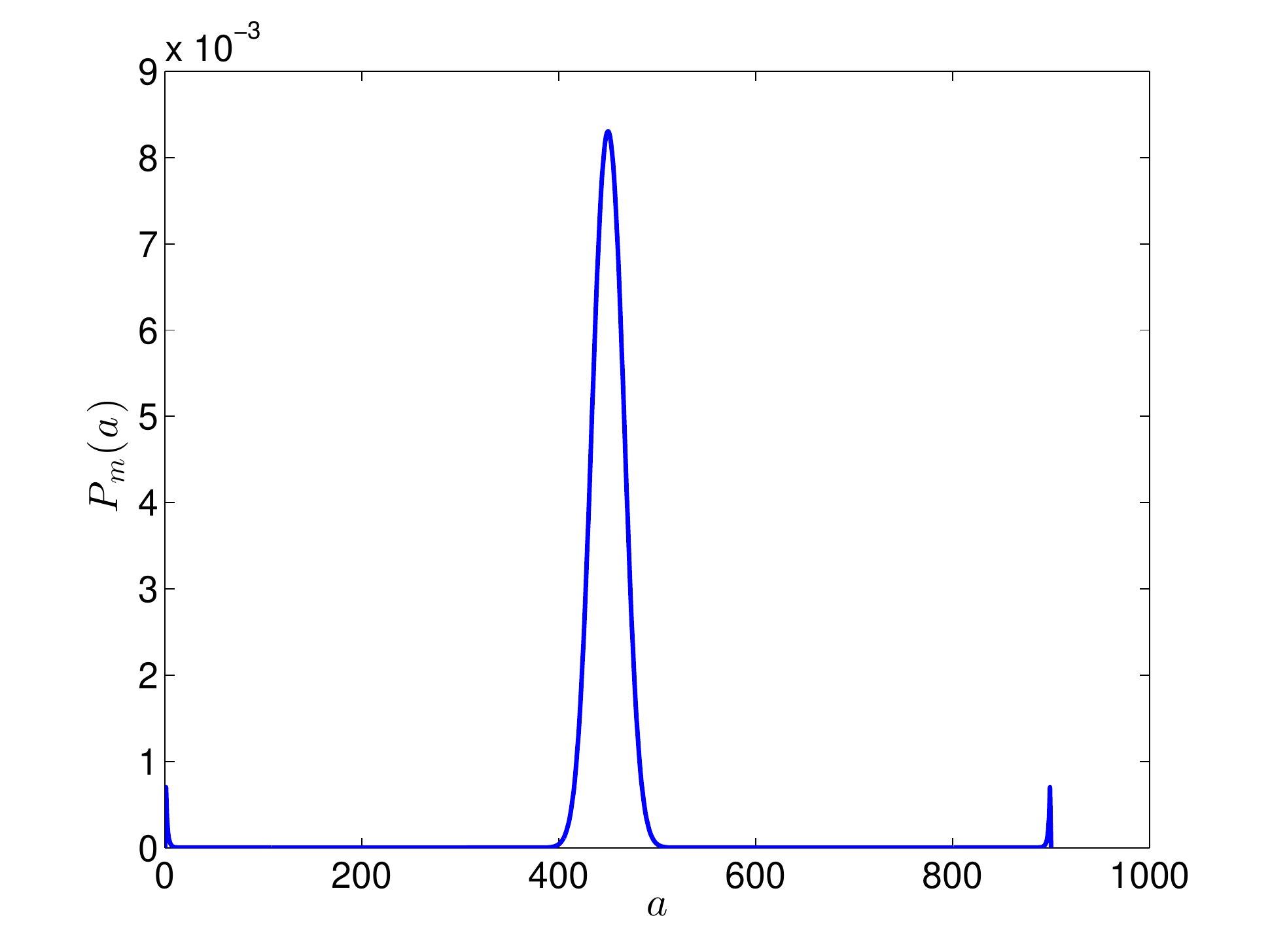}
\par\end{centering}

}
\par\end{centering}

\protect\caption{\label{fig:Pm_p_0.370}The size effect in $P_{m}$. $p=0.37$.}

\end{figure}

\par\end{center}

With this approximation (refer to supplementary material for details), 
\begin{align}
\begin{split}
&P_{3\rightarrow1}(T)\\
\approx&\left[1-C(N,p)\right]P_{3\rightarrow2,0}(T)\\
&+C(N,p)P_{2\rightarrow1|N/2}\left(T-\left\langle T_{3\rightarrow 2} \right\rangle\right),
\end{split}
\end{align}
where $\left\langle T_{3\rightarrow 2} \right\rangle$ is the average time of process \rom{1}, which only depends on $p$ as shown in Fig.~\ref{fig:avg_T_3to2}.
Now we can see the mechanism behind the acceleration brought about by having three opinions: there is a finite probability $[1-C(N,p)]/2$ that the population
will reach the consensus state with a time scale that is independent
of the population size, and otherwise at the beginning of process
\rom{2}, the population is a polarized state, where the distribution of the two surviving opinions is almost uniform, with a time scale that is an
increasing function of the population size.

The average time to consensus $\left\langle T_{3\rightarrow1}\right\rangle $ is (see the supplementary materials for the derivation):
\begin{align}
\left<T_{3\rightarrow1}\right> \approx \left< T_{3\rightarrow 2} \right>+C(N,p)\int_{0}^{\infty}\tau P_{2\rightarrow1|N/2}(\tau)d\tau.
\label{eq:avg_tttc}
\end{align}
According to Ref.~\cite{benczik_opinion_2009}, $\int_{0}^{\infty}\tau P_{2\rightarrow1|N/2}(\tau)d\tau\sim e^{a(p,M_{0})N}$.
Therefore, as $N\rightarrow\infty$, $\left<T_{3\rightarrow1}\right>$ is dominated by $e^{a(p,M_{0})N}$ and the acceleration of the time to consensus, if it exists, is a finite-size effect. 
The dependence of $1-C(N,p)$ on $N$ and $p$ is shown in Fig.~\ref{fig:b_N_p}.
When $p<0.5$, $1-C(N,p)$ decreases
exponentially with population size $N$. Since $1-C(N,p)$ is a decreasing
function of $N$, as $N\rightarrow\infty$, $\left< T_{3\rightarrow1}\right> \sim e^{a(p,M_{0})N}$,
and the acceleration is suppressed.
Fig.~\ref{fig:T2_T3_prediction_p0.370} shows the average
time to consensus $\left< T_{3\rightarrow1}\right>$ for two-state model
and three-state model when $p=0.37$, along with the predicted time
to consensus for three-state model calculated using Eq.~\ref{eq:avg_tttc}.
Note that the prediction fits $\left\langle T_{3\rightarrow1}\right\rangle $
very well.

\begin{center}
\begin{figure}
\begin{centering}
\includegraphics[width=0.9\columnwidth]{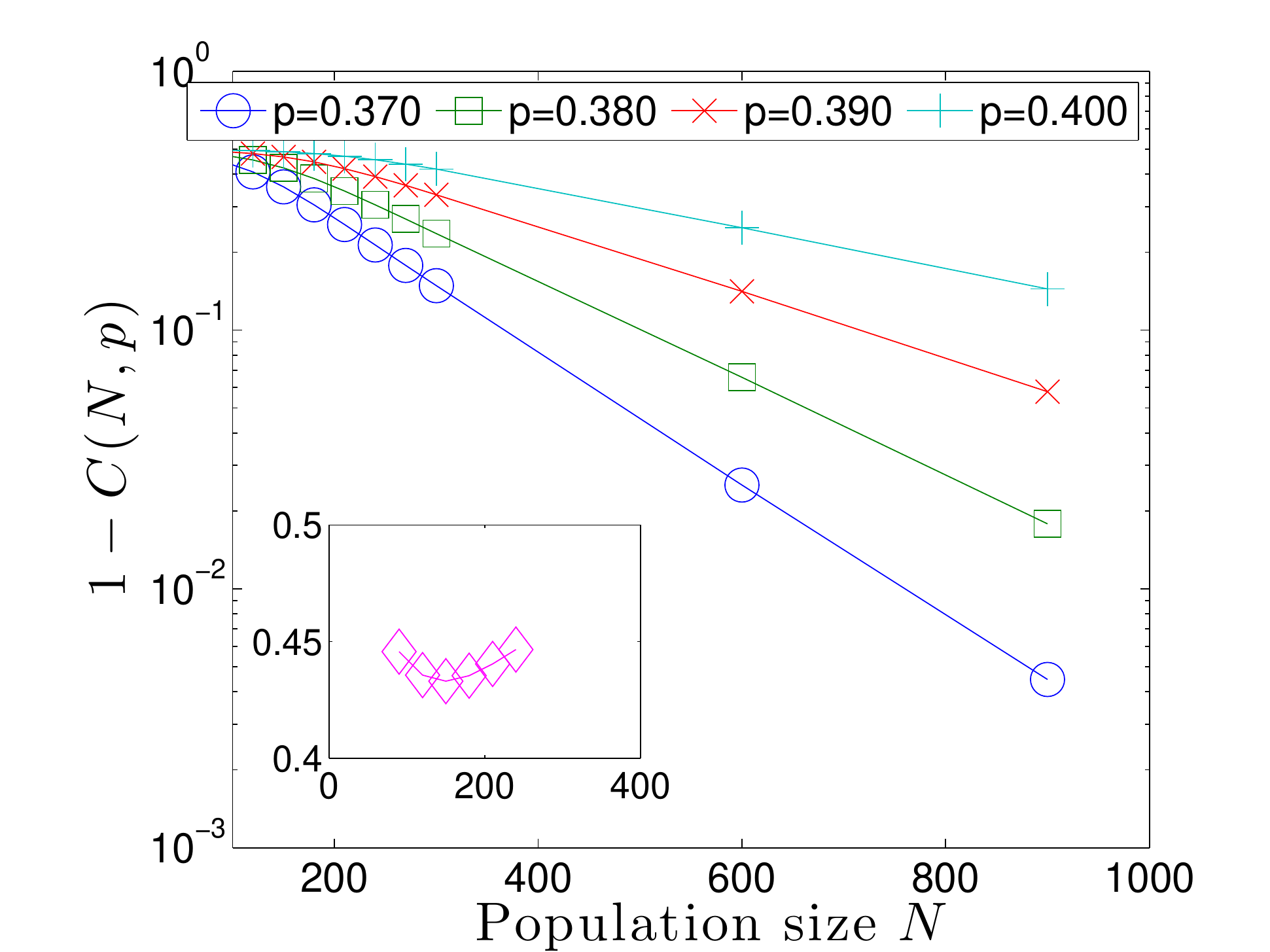}
\par\end{centering}

\protect\caption{\label{fig:b_N_p}$1-C(N,p)$ as a function of the population size for
various values of $p$. The inset shows $b(N,p)$ when $p=0.65$.}

\end{figure}

\par\end{center}

\begin{center}
\begin{figure}
\begin{centering}
\includegraphics[width=0.9\columnwidth]{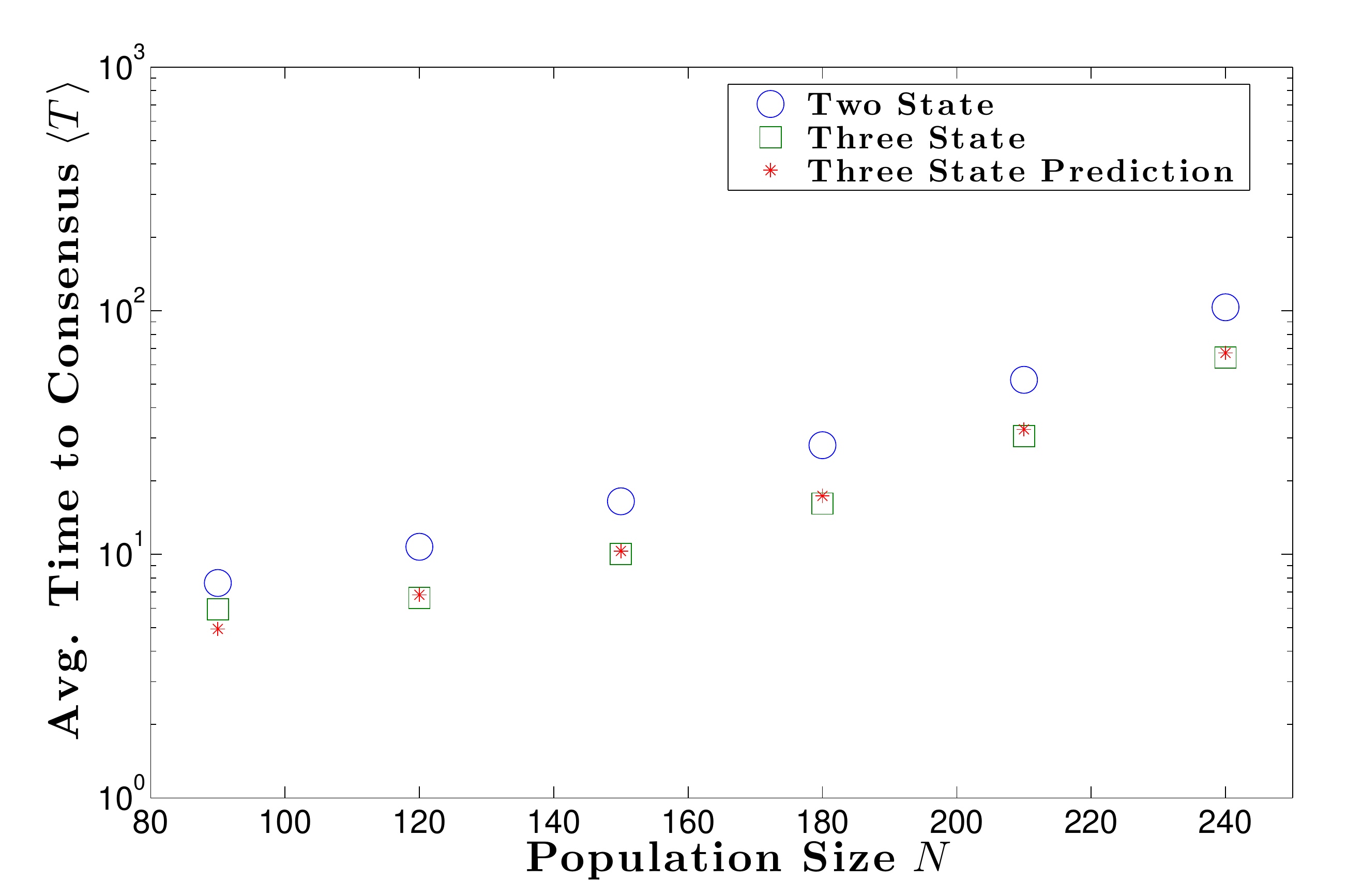}
\par\end{centering}

\protect\caption{\label{fig:T2_T3_prediction_p0.370}Average time to consensus for
two-state model, three-state model and that predicted by Eq.~\ref{eq:avg_tttc}.
$p=0.37$.}

\end{figure}

\par\end{center}

In the case with $p>0.5$, the approximation described by Eq.~\ref{eq:P_m_approx}
does not hold as good as in the case with $p<0.5$. (refer to the insert in Fig.~\ref{fig:b_N_p} for p>0.5) Numerical results
show that the following is a better approximation:

\begin{equation}
P_{m}(a)\approx\begin{cases}
b(N,p) & a=0,N\\
f(a) & m_{0}\le a\le m_{1}\\
0 & \text{otherwise}
\end{cases},\label{eq:P_m_approx_2}
\end{equation}
where $f(m)$ is symmetric with respect to $N/2$ and $\int_{m_{0}}^{m_{1}}f(a)da=1-2b(N,p)$.
Therefore,
\begin{equation}
\begin{aligned}
P(T_{3\rightarrow1})\approx & 2b(N,p)P_{3\rightarrow2,0}(T)\\
 & +\int\int_{m_{0}}^{m_{1}}P_{3\rightarrow2,a}\left(T-t\right)f(a)P_{2\rightarrow1|a}\left(t\right)da\, dt\\
\approx & 2b(N,p)P_{3\rightarrow2,0}(T)+\int_{m_{0}}^{m_{1}}f(a)P_{2\rightarrow1|a}\left(T-\bar{t}_{3\rightarrow2}\right)da
\end{aligned}
\end{equation}
and
\begin{equation}
\begin{aligned}
\left<T_{3\rightarrow1}\right>= & 2b\bar{t}_{3\rightarrow2|0}+\int_{\bar{t}_{3\rightarrow2}}^{\infty}\text{\ensuremath{\int}}_{m_{0}}^{m_{1}}f(a)T\, P_{2\rightarrow1|a}\left(T-\bar{t}_{3\rightarrow2}\right)da\, dT\\
= & 2b\bar{t}_{3\rightarrow2}+\text{\ensuremath{\int}}_{m_{0}}^{m_{1}}f(a)\left[\int_{0}^{\infty}\tau P_{2\rightarrow1|a}(\tau)d\tau+\bar{t}_{3\rightarrow2}\right]da\\
= & \bar{t}_{3\rightarrow2}+\text{\ensuremath{\int}}_{m_{0}}^{m_{1}}f(a)\int_{0}^{\infty}\tau P_{2\rightarrow1|a}(\tau)d\tau\, da\\
= & \bar{t}_{3\rightarrow2}+\text{\ensuremath{\int}}_{m_{0}}^{m_{1}}f(a)\left\langle T_{2\rightarrow1}\right\rangle _{a}\, da\\
\le & \bar{t}_{3\rightarrow2}+\text{\ensuremath{\int}}_{m_{0}}^{m_{1}}f(a)\left\langle T_{2\rightarrow1}\right\rangle _{N/2}\, da\\
= & \bar{t}_{3\rightarrow2}+\left[1-2b(N,p)\right]\left\langle T_{2\rightarrow1}\right\rangle _{N/2},
\end{aligned}
\end{equation}
where $\left\langle T_{2\rightarrow1}\right\rangle _{a}$ is effectively the time to consensus for two-state model given the initial condition is $m\equiv N_{\alpha}=a$, and the inequality follows from the fact that $\left\langle T_{2\rightarrow1}\right\rangle _{a}\le\left\langle T_{2\rightarrow1}\right\rangle _{N/2}$.
Therefore, the previous expression is effectively an upper bound for the time to consensus when $p>0.5$. 
See Fig.~\ref{fig:T2_T3_prediction_p0.650} for $\left\langle T_{3\rightarrow1}\right\rangle $ when $p=0.65$. 
The value predicted by Eq.~\ref{eq:avg_tttc} consistently serves as the upper bound for $\left\langle T_{3\rightarrow1}\right\rangle $. 
When $p=0.65$ , $\left<T_{3\rightarrow2}\right>\sim\exp(0.026N)$
while $\left<T_{2\rightarrow1}\right>\sim\exp(0.04N)$. The acceleration when
$p>0.5$ is a combination of two acceleration effects: 1) with probability
$2b(N,p)$, $\left<T_{3\rightarrow1}\right>$ is dominated by $\left<T_{3\rightarrow2}\right>$,
which although depends on the population size exponentially, the
exponential coefficient is significantly smaller than that of $\left<T_{2\rightarrow1}\right>$.
2) with probability $1-2b(N,p)$, the population reaches the consensus
state in a time scale at most $\left<T_{3\rightarrow2}\right>+\left\langle T_{2\rightarrow1}\right\rangle \approx\left\langle T_{2\rightarrow1}\right\rangle $
in large $N$ limit.

\begin{center}
\begin{figure}
\begin{centering}
\includegraphics[width=0.9\columnwidth]{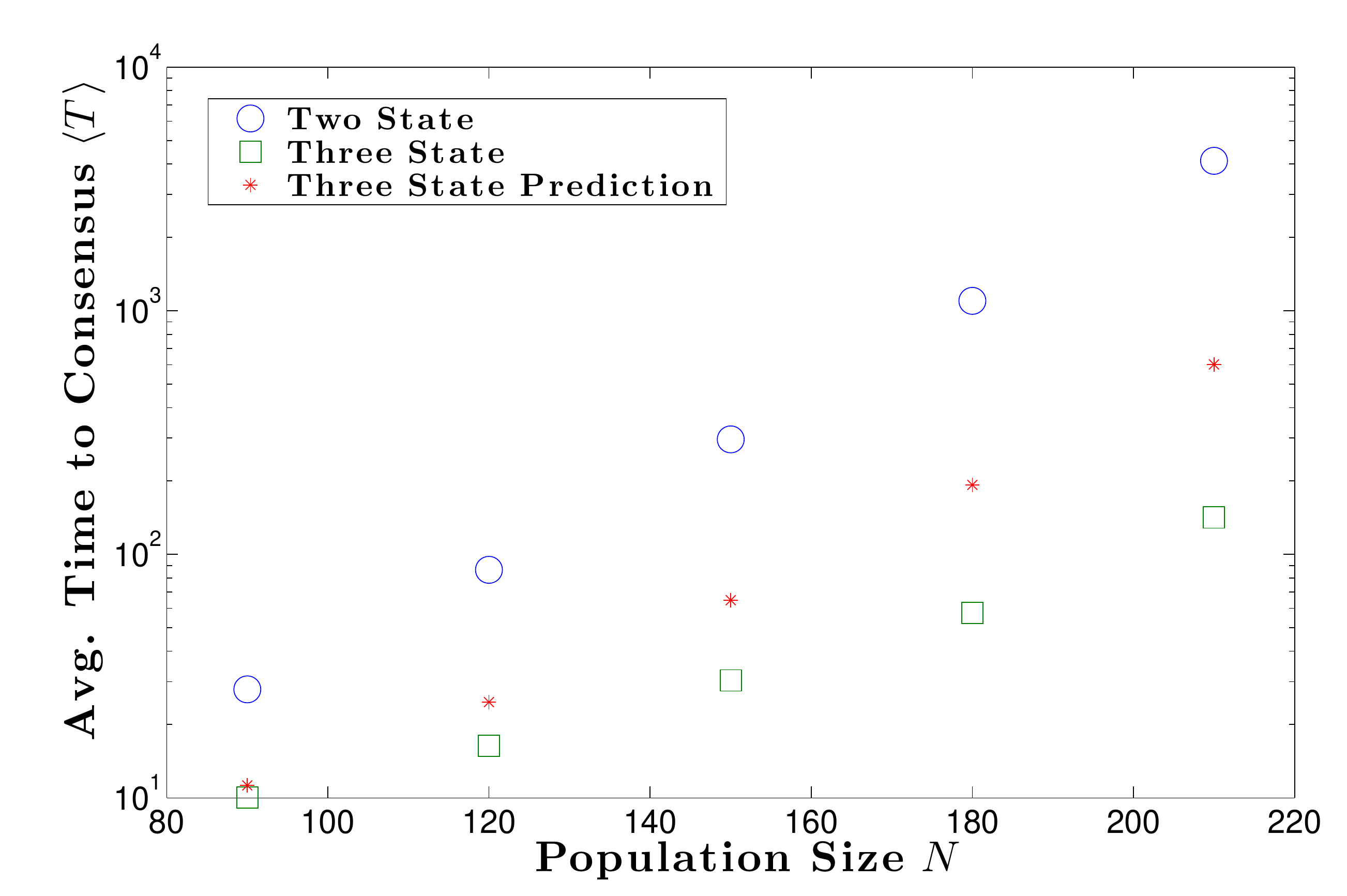}
\par\end{centering}

\protect\caption{\label{fig:T2_T3_prediction_p0.650}Average time to consensus for
two-state model, three-state model and that predicted by Eq.~\ref{eq:avg_tttc}.
$p=0.65$. $\left\langle T_{2\rightarrow1}\right\rangle \sim\exp(0.04N)$.}

\end{figure}

\par\end{center}

\section{Conclusions\label{sec:Conclusions}}

In this work, we have proposed a generalization of the opinion formation
model proposed in Ref.~\cite{benczik_opinion_2009}. The proposed
model is a plurality vote model on random adaptive networks. Through
numerical simulation, we have shown that when $p$ is smaller than
or larger than 0.5, the times to consensus of more-than-two-state
models are statistically shorter than the that of two-state model.
To understand the mechanism behind the acceleration in time to consensus
induced by having more than two opinions, we have broken up the whole
process of three-state population reaching consensus state into two
subprocesses: process \rom{1} is where one of the three initial available opinions goes extinct and process \rom{2} is where one of the two remaining opinions at the end of process \rom{1} goes extinct.
 For $p<0.5$, the time scale of process \rom{1} is
independent of the population size. The time scale of process \rom{2}
can be vastly different depending on the state of the population at
the end of process \rom{1}. The population can either be at the vicinity
of the consensus state, such that reaching the consensus state is
instantaneous, or in the polarized state such that the relaxation
into the consensus state has a very long time scale. For $p>0.5$,
the average time of process \rom{1}, $\left<T_{3\rightarrow2}\right>$,
depends on the size of the population exponentially, but with an exponential
coefficient significantly smaller than the counter-part in $\left<T_{2}\right>$.
The time scale of process \rom{2} follows the same mechanism as when $p>0.5$. This
results in a remarkable feature in our plurality vote model on adaptive
networks: 1) when $p<0.5$, there is a non-zero probability, decay
exponentially with the size of the population, that the population
reach to consensus state in a time scale that is independent of the
population size. 2) when $p>0.5$, there is a finite probability $2b(N,p)$
that the population reaches the consensus state in time scale which
is dependent on the population size exponentially but has an exponential
coefficient significantly smaller than $\left<T_{2}\right>$, and
there is $1-2b(N,p)$ probability that the population reaches the
consensus state in a time scale that is comparable to but could be
smaller than $\left<T_{2}\right>$. From the combination of numerical
and analytical analysis, having more available opinions does not mean
it is harder, or takes longer to reach the consensus state. It remains
an open question whether the acceleration in the time to consensus
is due to a similar mechanism for the plurality model that has more
than three opinions.

\putbib[mylib]
\end{bibunit}

\pagebreak
\widetext
\begin{center}
\textbf{\large Supplemental Materials: Does Having More Options Mean Harder to Reach Consensus?}
\end{center}

\setcounter{equation}{0}
\setcounter{figure}{0}
\setcounter{table}{0}
\setcounter{page}{1}
\makeatletter
\renewcommand{\theequation}{S\arabic{equation}}
\renewcommand{\thefigure}{S\arabic{figure}}
\renewcommand{\bibnumfmt}[1]{[S#1]}
\renewcommand{\citenumfont}[1]{S#1}

\makeatother

\begin{bibunit}[unsrt]

\author{Degang Wu, Kwok Yip Szeto}
\title{Supplementary Materials}
\maketitle
\section{Alternative form of the master equation}
Because of the symmetry in the transition probability, $W_{1\rightarrow3}(N_1,N_2)$, $W_{2\rightarrow1}(N_1,N_2)$ can be obtained by simple transformation of $W_{1\rightarrow2}(N_1,N_2)$. The following form of master equation only contains $W_{1\rightarrow2}$ and hence is computational-efficient:
\begin{align}
\begin{split}
&\partial_{t}P(N_1,N_2)\\
=&-\left[W_{1\rightarrow2}(N_1,N_2)+W_{1\rightarrow2}(N_1,N_3)+W_{1\rightarrow2}(N_2,N_1)+W_{1\rightarrow2}(N_2,N_3)+\right.\\
&\left.W_{1\rightarrow2}(N_3,N_1)+W_{1\rightarrow2}(N_3,N_2)\right]P(N_1,N_2)\\
 & +W_{1\rightarrow2}(N_1+1,N_2-1)P(N_1+1,N_2-1)+W_{1\rightarrow2}(N_1+1,N_3-1)P(N_1+1,N_2)\\
 &+W_{1\rightarrow2}(N_2+1,N_1-1)P(N_1-1,N_2+1)+W_{1\rightarrow2}(N_2+1,N_3-1)P(N_1,N_2+1)\\
 &+W_{1\rightarrow2}(N_3+1,N_1-1)P(N_1-1,N_2)+W_{1\rightarrow2}(N_3+1,N_2-1)P(N_1,N_2-1).
\end{split}
\end{align}
\section{Approximating $W_{1\rightarrow2}$ in thermodynamics limit}
The transition probabilities for a chosen agent to hold opinion 1 and subsequently changes to hold opinion 2, denoted as $W_{1\rightarrow2}$, is defined mathematically as
\begin{equation}
W_{1\rightarrow2}(N_1,N_2)= \frac{N_1}{N}\sum_{l=0}^{N_1-1}\sum_{l'=0}^{N_2}\sum_{l''=0}^{N_3}B_{N_1-1,p}(l)B_{N_2,q}(l')B_{N_3,q}(l'')
 \Theta(l'-l)\Theta(l'-l''),
\end{equation}
where $\Theta(l)$ is the step function and $B_{n,p}(l)={n \choose l}p^l(1-p)^{n-l}$, the probability mass function of a binomial distribution $B(n,p)$. As $n\rightarrow\infty$, $B(n,p)$ asymptotically approaches the Gaussian distribution $\mathcal{N}(np,np(1-p))$. Therefore, $B(N_1/N,p)$ approaches $\mathcal{N}(p N_1/N,p(1-p) N_1/N^2)$ as $N_1,N\rightarrow\infty$, $p$ and $x=N_1/N$ held fixed and the spread of the distribution becomes narrower as $N$ becomes larger. 

First, let us investigate $w_{1\rightarrow2}\equiv \frac{N}{N_1}W_{1\rightarrow2}$. Rearrange the terms such that
\begin{equation}
w_{1\rightarrow2}(N_1,N_2)= \sum_{l'=0}^{N_2}B_{N_2,q}(l')\left[\sum_{l=0}^{N_1-1}\sum_{l''=0}^{N_3}B_{N_1-1,p}(l)B_{N_3,q}(l'')
 \Theta(l'-l)\Theta(l'-l'')\right].
\end{equation}
With changes of variables $x=N_1/N,y=N_2/N,z=N_3/N$, the terms in the square brackets can be approximated as an integral:
\begin{equation}
I(x,z,s')=\int_0^{x-\epsilon}\int_0^{z}\dfrac{1}{\sqrt{2\pi}\sigma_x}e^{-\frac{(s-(x-\epsilon)p)^2}{\sigma_x^2}}\dfrac{1}{\sqrt{2\pi}\sigma_x^2}e^{-\frac{(s''-zq)^2}{\sigma_x^2}}\Theta(s'-s)\Theta(s'-s'')dsds'',
\end{equation}
where $\epsilon=1/N$ and $s=l/N$. It is clear that if $s'<(x-\epsilon)p$ or $s'<zq$, the double integral is zero. $w_{1\rightarrow2}$ then can be approximated by
\begin{equation}
w_{1\rightarrow2}\approx\int_0^y I(x,z,s')\dfrac{1}{\sqrt{2\pi}\sigma_y}e^{-\frac{(s'-yq)^2}{\sigma_y^2}}ds'.
\end{equation}
Since most of the probability density concentrate near $yq$ in the thermodynamics limit, $w_{1\rightarrow2}$ is non-zero only when $yq>(x-\epsilon)p$ and $yq>zq$ simultaneously. 

Therefore, in the thermodynamics limit, $w_{1\rightarrow2}=1$ when $y>\max(p(x-\epsilon)/q,(N-x)/2)$ and zero otherwise. In other words, $W_{1\rightarrow2}=x/N$ when $y>\max(p(x-\epsilon)/q,(N-x)/2)$ and zero otherwise. Approximation of other transition probabilities can be made in similar ways.

\section{Deriving master equation in region 1 of $W_{1\rightarrow2}$}

In region 1 of $W_{1\rightarrow2}$,
or $\mathbb{A}\left[(1,2),(3,2)\right]$, $W_{1\rightarrow2}$ and $W_{3\rightarrow2}$
are positive and proportional to $x$ and $z$, respectively, and other transition probabilities are zero.
Therefore, the master equation can be approximated by:
\begin{align}
\begin{split}
&\partial_{t}P(N_1,N_2)\\
\approx & -\left[W_{1\rightarrow2}(N_1,N_2)+W_{3\rightarrow2}(N_1,N_2)\right]P(N_1,N_2)+W_{1\rightarrow2}(N_1+1,N_2-1)P(N_1+1,N_2-1)\\
&+W_{3\rightarrow2}(N_1,N_2-1)P(N_1,N_2-1)\\
= & -\left[W_{1\rightarrow2}(N_1,N_2)+W_{1\rightarrow2}(N_3,N_2)\right]P(N_1,N_2)+W_{1\rightarrow2}(N_1+1,N_2-1)P(N_1+1,N_2-1)\\
&+W_{1\rightarrow2}(N_3+1,N_2-1)P(N_1,N_2-1)\\
\approx & -\left(\frac{N_1+N_3}{N}\right)P(N_1,N_2)+\left(\frac{N_1+1}{N}\right)P(N_1+1,N_2-1)+\left(\frac{N_3+1}{N}\right)P(N_1,N_2-1).
\end{split}
\end{align}
Consider the limit $N\rightarrow\infty$, and normalize $N_1,N_2,N_3$ by
$x=N_1/N$, etc., the master equation can be further approximated
by
\begin{align}
\begin{split}
\partial_{t}P(x,y)= & -\left[x+z\right]P(x,y)+(x+\epsilon)P(x+\epsilon,y-\epsilon)+(z+\epsilon)P(x,y-\epsilon)\\
= & -\left[x+z\right]P(x,y)+(x+\epsilon)\left[P(x,y)+\frac{\partial P}{\partial x}\epsilon+\frac{\partial P}{\partial y}(-\epsilon)+\cdots\right]\\
&+(z+\epsilon)\left[P(x,y)+\frac{\partial P}{\partial y}(-\epsilon)+\cdots\right]\\
\approx & \epsilon\left[x\frac{\partial P}{\partial x}+(y-1)\frac{\partial P}{\partial y}+2P(x,y)\right],
\end{split}
\end{align}
where $\epsilon=1/N$ and the last approximation keeps
only the 1st order terms of $\epsilon$.

\section{Deriving master equation in region 4 of $W_{1\rightarrow2}$}

Similarly, in region 4, or $\mathbb{A}\left[(3,2),(1,2)\right]$,
only $W_{1\rightarrow2},W_{2\rightarrow3}$ and $W_{3\rightarrow2}$ are positive, and the master equation
in that region can be approximated by, in the $N\rightarrow\infty$
limit, 
\begin{align}
\begin{split}
&\partial_{t}P(x,y)\\
= & -\left[x+y+z\right]P(x,y)+(x+\epsilon)P(x+\epsilon,y-\epsilon)+(y+\epsilon)P(x,y+\epsilon)\\
&+(z+\epsilon)P(x,y-\epsilon)\\
\approx & \epsilon\left[2P(x,y)+x\frac{\partial P}{\partial x}+(2y-1)\frac{\partial P}{\partial y}\right].
\end{split}
\end{align}

\section{Deriving $P_{3\rightarrow1}(T)$}
Since
\begin{equation}
P_{m}(a)\approx\begin{cases}
(1-C(N,p))/2 & a=0,N\\
C(N,p) & a=N/2\\
0 & \text{otherwise}
\end{cases},\label{eq:P_m_approx}
\end{equation}
\begin{align}
\begin{split}
&P_{3\rightarrow1}(T)\\
= & \int_{0}^{N} da \int_{0}^{T}P_{3\rightarrow2,2\rightarrow1}(T-t,t|a)P_{m}(a)dt\\
= & \int_0^Nda\int_0^T P_{3\rightarrow2,a}(T-t)P_{2\rightarrow1|a}(t)P_{m}(a)dt\\
\approx & \int P_{3\rightarrow2,0}(T-t)P_{m}(0)P_{2\rightarrow1|0}(t)dt+\int P_{3\rightarrow2,N}(T-t)P_{m}(N)P_{2\rightarrow1|N}(t)dt\\
 & +\int P_{3\rightarrow2,\frac{N}{2}}\left(T-t\right)P_{m}\left(\frac{N}{2}\right)P_{2\rightarrow1|\frac{N}{2}}\left(t\right)dt\\
= & \int P_{3\rightarrow2,0}(T-t) \left[1-C(N,p)\right] P_{2\rightarrow1|0}(t)dt+\int P_{3\rightarrow2,\frac{N}{2}}\left(T-t\right)C(N,p)P_{2\rightarrow1|\frac{N}{2}}\left(t\right)dt\\
\approx & \left[1-C(N,p)\right]P_{3\rightarrow2,0}(T)+C(N,p)P_{2\rightarrow1|N/2}\left(T-\left< T_{3\rightarrow 2} \right>_{N/2}\right)\\
\approx & \left[1-C(N,p)\right]P_{3\rightarrow2,0}(T)+C(N,p)P_{2\rightarrow1|N/2}\left(T-\left< T_{3\rightarrow 2} \right>\right),
\end{split}
\end{align}
where the fifth (approximate) equality is due to the fact that compared
to $P_{2\rightarrow1}(T)$, $P_{3\rightarrow2}(T)$ is narrowly peaked
at a value $\left< T_{3\rightarrow 2}\right>_{N/2}$ and hence it can be regarded
as a delta function centered at $T=\left< T_{3\rightarrow 2}\right>_{N/2}$, which is the average time of process \rom{1} given that at the end of process \rom{1}, $N_\alpha=N/2$. The final (approximate) equality is due to that $\left< T_{3\rightarrow 2}\right>_{N/2} \approx \left< T_{3\rightarrow 2}\right>_0 \approx \left< T_{3\rightarrow 2} \right>$.

\section{Deriving $\left<T_{3\rightarrow1}\right>$}

\begin{align}
\begin{split}
&\left<T_{3\rightarrow1}\right>\\
=&\int_0^\infty \tau P(\tau) d\tau\\
= & [1-C(N,p)]\int_{0}^{\infty}\tau P_{3\rightarrow2,0}(\tau)d\tau+C(N,p)\int_{\left< T_{3\rightarrow 2} \right>}^{\infty}\tau\, P_{2\rightarrow1|N/2}\left(\tau-\left< T_{3\rightarrow 2} \right>\right)d\tau \\
\approx & [1-C(N,p)]\left< T_{3\rightarrow 2} \right>+C(N,p)\int_{0}^{\infty}\left(\tau+\left< T_{3\rightarrow 2} \right>\right)P_{2\rightarrow1|N/2}(\tau)d\tau \\
\approx & [1-C(N,p)]\left< T_{3\rightarrow 2} \right>+C(N,p)\left[\int_{0}^{\infty}\tau P_{2\rightarrow1|N/2}(\tau)d\tau+\left< T_{3\rightarrow 2} \right>\right] \\
= & \left< T_{3\rightarrow 2} \right>+C(N,p)\int_{0}^{\infty}\tau P_{2\rightarrow1|N/2}(\tau)d\tau.
\end{split}
\end{align}

\section{Deriving $\left<T_{2\rightarrow1}\right>_{N/2}\ge\left<T_{2\rightarrow1}\right>_a$}
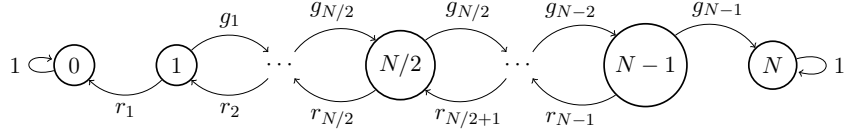
\begin{figure}
\centering
{\scalebox{0.8}{
\begin{tikzpicture}[->,start chain,regular/.style={draw,thick,circle},bend angle=45]
	\node [on chain, regular] (0) {0};
	\node [on chain, regular] (1) {1};
	\node [on chain] (2) {$\cdots$};
	
	\node [on chain, regular] (N/2) {$N/2$};
	
	\node [on chain] (N-2) {$\cdots$};
	\node [on chain, regular] (N-1) {$N-1$};
	\node [on chain, regular] (N) {$N$};
	
	\path
		(1) edge [bend left] node [below] {$r_1$} (0)
		(0) edge [loop left] node [left] {1} (0)
		(1) edge [bend left] node [above] {$g_1$} (2)
		(2) edge [bend left] node [below] {$r_2$} (1)
		(2) edge [bend left] node [above] {$g_{N/2}$} (N/2)
		(N/2) edge [bend left] node [below] {$r_{N/2}$} (2)
		(N/2) edge [bend left] node [above] {$g_{N/2}$} (N-2)
		(N-2) edge [bend left] node [below] {$r_{N/2+1}$} (N/2)
		(N-2) edge [bend left] node [above] {$g_{N-2}$} (N-1)
		(N-1) edge [bend left] node [above] {$g_{N-1}$} (N)
		(N-1) edge [bend left] node [below] {$r_{N-1}$} (N-2)
		(N) edge [loop right] node [right] {1} (N);
\end{tikzpicture}}}
\caption{The evolution of a population with binary opinion on adaptive networks~\cite{benczik_opinion_2009} can be regarded as a Markov chain process $(X_k)$ with two absorbing states. $g_x$ is the transition probability state $x$ to state $x+1$ while $r_x$ is the transition probability from state $x$ to $x-1$. The explicit expression of $g_x$ and $r_x$ can be found in Ref.~\cite{benczik_opinion_2009}, but the proof here does not need the explicit form of $g_x$ and $r_x$ except that $g_x=r_{N-x}$.}
\label{fig:original_markov_chain}
\end{figure}

Here we provide the derivation that $\left<T_{2\rightarrow1}\right>_{N/2}\ge\left<T_{2\rightarrow1}\right>_a$. The evolution of a population with binary opinion on adaptive networks~\cite{benczik_opinion_2009} can be regarded as a Markov chain process $(X_k)$ depicted in Fig.~\ref{fig:original_markov_chain}, where the transition probabilities satisfy $g_x=r_{N-x}$. For simplicity, we assume the population size $N$ is even. 

We denote by $\pi_{0,n}$ the probability that the chain enters absorbing state 0 when the initial state is $n$, and $\pi_{N,n}$ is similarly defined. Denote by $\tau_{0,n}$ the first passage time of state 0 when the initial state is $n$ and $\tau_{N,n}$ is similarly defined. The mean time that the chain starting at state $n$ hits either one of the two absorbing states is, therefore, $\tau_{n}=\pi_{0,n}\tau_{0,n}+\pi_{N,n}\tau_{N,n}$. By definition, $\tau_{a}\equiv\left<T_{2\rightarrow1}\right>_a$. To prove that $\left<T_{2\rightarrow1}\right>_a\le\left<T_{2\rightarrow1}\right>_{N/2}$ is equivalent to proving $\tau_{a}\le\tau_{N/2}$.

We consider a \textit{lumped} Markov chain~\cite{snell_finite_1969} $(Y_k)$ which is defined by
\begin{equation}
Y_k=u(X_k),\qquad u(x)=\max\{x,N-x\},
\end{equation}
which partitions into $\{A_N,A_{N-1},\cdots,A_{N/2}\}$, where $A_N=\{0,N\},A_{N-1}=\{1,N-1\},\cdots,A_{N/2}=\{N/2\}$.

\begin{figure}
\centering
{\scalebox{0.8}{
\begin{tikzpicture}[->,start chain,regular/.style={draw,thick,circle},bend angle=45]	
	\node [on chain, regular] (N/2) {$N/2$};
	\node [on chain, regular] (N/2+1) {$N/2+1$};
	\node [on chain] (N-2) {$\cdots$};
	\node [on chain, regular] (N-1) {$N-1$};
	\node [on chain, regular] (N) {$N$};
	
	\path
		(N/2) edge [bend left] node [above] {$2g_{N/2}$} (N/2+1)
		(N/2+1) edge [bend left] node [below] {$r_{N/2+1}$} (N/2)
		(N/2+1) edge [bend left] node [above] {$g_{N/2+1}$} (N-2)
		(N-2) edge [bend left] node [below] {$r_{N/2+2}$} (N/2+1)
		(N-2) edge [bend left] node [above] {$g_{N-2}$} (N-1)
		(N-1) edge [bend left] node [above] {$g_{N-1}$} (N)
		(N-1) edge [bend left] node [below] {$r_{N-1}$} (N-2)
		(N) edge [loop right] node [right] {1} (N);
\end{tikzpicture}}}
\caption{The lumped Markov chain $(Y_k)$.}
\label{fig:lumped_markov_chain}
\end{figure}

According to Ref.~\cite{snell_finite_1969}, the lumped process $(Y_k)$ is a Markov chain if and only if for every pair of sets $A_i$ and $A_j$, $\sum_{l\in A_j}p_{kl}$ has the same value for every $k$ in $A_i$. This is indeed true in our case because of the symmetry in transition probabilities $g_x=r_{N-x}$. The lumped Markov chain is described in Fig.~\ref{fig:lumped_markov_chain}.

Now consider the process $(Y_k)$. The mean first passage time $\tau_{N,u(n)}$ for $(Y_k)$ is the same as the mean first passage time $\tau_{n}$ for $(X_k)$. According to Ref.~\cite{van_kampen_stochastic_1992}, $\tau_{N,N/2}>\tau_{N,N/2+1}>\cdots>\tau_{N,N-1}$. Therefore, for the original Markov chain $X_k$, $\tau_{1}<\tau_{2}<\cdots<\tau_{N/2-1}<\tau_{N/2}>\tau_{N/2+1}>\cdots>\tau_{N-1}$. In the case where $N$ is odd, we can reach the same conclusion in similar way.

Therefore, we have proved that $\left<T_{2\rightarrow1}\right>_a\le\left<T_{2\rightarrow1}\right>_{N/2}$.

\putbib[mylib]
\end{bibunit}

\end{document}